\documentclass[twocolumn, usenatbib ]{aastex62}
\usepackage{graphicx}
\def\ltsima{$\; \buildrel < \over \sim \;$}
\def\simlt{\lower.5ex\hbox{\ltsima}}
\def\gtsima{$\; \buildrel > \over \sim \;$}
\def\simgt{\lower.5ex\hbox{\gtsima}}
\newcommand {\sfr}{M$_{\odot}$\,yr$^{-1}$}

\newcommand {\um}{$\mu$m}

\newcommand {\msun}{M$_{\odot}$}

\newcommand {\gt}{$>$}

\newcommand {\herschel}{\textit{Herschel}}

\newcommand {\halpha}{H$\alpha$}
\newcommand {\magphys}{\textsc{magphys}}
\newcommand {\kms}{km\,s$^{-1}$}

\def\ltsima{$\; \buildrel < \over \sim \;$}
\def\simlt{\lower.5ex\hbox{\ltsima}}
\def\gtsima{$\; \buildrel > \over \sim \;$}
\def\simgt{\lower.5ex\hbox{\gtsima}}

\newcommand {\scubaii}{{\sc Scuba-2}}

\newcommand {\alphaco}{$\alpha_{\rm CO}$}
\newcommand {\acounits}{M$_{\odot}$/(K\,km\,s$^{-1}$\,pc$^{2}$)}

\shorttitle{Gas Characterization in a z=2.5 Protocluster}

\begin{document}

\title{Comprehensive Gas Characterization of a $z=2.5$ Protocluster: A Cluster Core Caught in the Beginning of Virialization?}

\author[0000-0002-6184-9097]{Jaclyn~B.~Champagne}
\author[0000-0002-0930-6466]{Caitlin~M.~Casey}
\author[0000-0002-7051-1100]{Jorge~A.~Zavala}
\affil{Department of Astronomy, University of Texas at Austin, 2515 Speedway, Stop C1400, Austin, TX 78751, USA}

\author[0000-0002-3892-0190]{Asantha~Cooray}
\affil{Department of Physics and Astronomy, University of California, Irvine, CA 92697, USA}

\author[0000-0001-7147-3575]{Helmut~Dannerbauer}
\affil{Instituto de Astrof\'{i}sica de Canarias (IAC), E-38205 La Laguna, Tenerife, Spain}
\affil{Universidad de La Laguna, Dpto. Astrof\'{i}sica, E-38206 La Laguna, Tenerife, Spain}

\author[0000-0002-9378-4072]{Andrew~Fabian}
\affil{Institute of Astronomy, Madingley Road, Cambridge CB3 0HA}

\author[0000-0003-4073-3236]{Christopher~C.~Hayward}
\affil{Center for Computational Astrophysics, Flatiron Institute, 162 Fifth Avenue, New York, NY 10010, USA}

\author[0000-0002-7530-8857]{Arianna~S.~Long}
\affil{Department of Physics and Astronomy, University of California, Irvine, CA 92697, USA}

\author[0000-0003-3256-5615]{Justin~S.~Spilker}
\affil{University of Texas at Austin, 2515 Speedway, Stop C1400, Austin, TX 78751, USA}
\affil{NHFP Hubble Fellow}

\begin{abstract}

In order to connect galaxy clusters to their progenitor protoclusters, we must constrain the star formation histories within their member galaxies and the timescale of virial collapse. 
In this paper we characterize the complex star-forming properties of a $z=2.5$ protocluster in the COSMOS field using ALMA dust continuum and new VLA CO(1-0) observations of two filaments associated with the structure, sometimes referred to as the ``Hyperion" protocluster. 
We focus in particular on the protocluster ``core" which has previously been suggested as the highest redshift bona fide galaxy cluster traced by extended X-ray emission in a stacked \textit{Chandra/XMM} image.  
We re-analyze this data and refute these claims, finding that at least 40 $\pm$ 17\% of extended X-ray sources of similar luminosity and size at this redshift arise instead from Inverse Compton scattering off recently extinguished radio galaxies rather than intracluster medium.  
Using ancillary COSMOS data, we also constrain the SEDs of the two filaments' eight constituent galaxies from the rest-frame UV to radio.  
We do not find evidence for enhanced star formation efficiency in the core and conclude that the constituent galaxies are already massive (M$_{\star} \approx 10^{11}$\,\msun), with molecular gas reservoirs $>$10$^{10}$\,\msun\, that will be depleted within 200--400 Myr.   
Finally, we calculate the halo mass of the nested core at $z=2.5$ and conclude that it will collapse into a cluster of 2--9\,$\times\,10^{14}$\,\msun, comparable to the size of the Coma cluster at $z=0$ and accounting for at least 50\% of the total estimated halo mass of the extended ``Hyperion" structure.

\end{abstract}


\section{Introduction} \label{sec:intro}

The assembly of the most massive gravitationally bound structures at different epochs in the Universe's history is relatively well constrained theoretically, but observationally we are missing key pieces to fill in the precise timeline of their structure formation. 
Structures within a $\Lambda$CDM Universe form hierarchically, as demonstrated in cosmological simulations: the most massive structures in the Universe ($M_h>10^{14}$\,M$_{\odot}$) have collapsed into single halos only in the last 8--9 Gyr \citep{Springel2005a, Vogelsberger2014a}. 

Observations at very high redshifts ($z>5$) suggest that overdensities of massive, dust-rich galaxies may exist around rare, radio-loud quasars at very early times, demonstrating accelerated growth of galaxies during collapse \citep[e.g.,][]{Venemans2007c, Miley2008a, Decarli2017a} in line with the predictions of cosmic downsizing \citep{Cowie1986a}. 
However, other studies have found a lack of evidence for quasar-centered overdensities \citep[e.g.,][]{Farina2017a, Mazzucchelli2017b, Champagne2018a}, suggesting that quasars are not the only pathway for the formation of galaxy clusters at high-redshift, but instead one of many types of overdensities.

At lower redshifts ($z\leq1.5$), fully-formed galaxy clusters with halo masses $M_h\geq10^{14}$\,$\rm M_{\odot}$ are straightforward to identify because their observational characteristics are somewhat uniform \citep[see review in][]{Kravtsov2012a}. 
They are typically characterized by populations of massive elliptical galaxies following a tight red sequence \citep{Lewis2002a, Skibba2009a}, which is likely the result of suppressed star formation rates (SFR) in the process of cluster quenching \citep{Cooper2008a}. 
In addition to the ``red and dead'' galaxy population, galaxy clusters are almost completely virialized \citep{Xu2000a} and have reservoirs of hot gas in the intracluster medium (ICM), easily traced by strong, diffuse Bremsstrahlung in the X-ray \citep[e.g.,][]{Rosati2002a, Kravtsov2012a} and a Sunyaev-Zel'dovich signature in the millimeter due to absorption of Cosmic Microwave Background (CMB) photons \citep[e.g.,][]{Menanteau2012a, Mantz2014a, Bleem2015a}. 

What remains unconstrained are the intermediate physical processes by which cluster progenitors --- so-called protoclusters --- collapse and build up their mass in the intervening time of $2<z<5$. 
Protoclusters \citep{Overzier2016a} are structures expected to collapse into $\sim10^{14-15}$\,\msun\, clusters by the time they virialize before $z=0$ \citep{Chiang2013a}. 
Prior to their gravitational collapse, protoclusters began as spatially extended, large-scale overdensities along cosmic filaments \citep{Muldrew2015a}, occupying a fractional cosmic volume that shrinks by three orders of magnitude from $z=7$ to $z=0$ \citep{Chiang2017a}. 
Observationally, they can span several comoving Mpc at $z>5$, and have been shown to span half a degree on the sky at $z\sim2-3$, with comoving volumes $\geq$15,000\,Mpc$^3$  \citep{Matsuda2005a, Casey2016a, Hung2016a}. 
In terms of constituent galaxy evolution, the cosmic star formation rate density sees its peak at $2<z<4$ \citep{Madau2014a}, and galaxies in overdense regions are believed to experience the bulk of their star formation at $z>2$ \citep[e.g.,]{Collins2009a, Papovich201a}. 
Protoclusters observed across large fields are thus ideal laboratories for investigating the rapid buildup of stellar mass in overdense regions at cosmic noon.

One of the most important physical questions about these dense environments is what mode of star formation one would expect to see in galaxies within protoclusters at $z>2$ and whether this occurs through in-situ formation or accretion. 
If galaxies in these overdensities primarily build up their mass through short-lived bursts of star formation, consistent with rare dusty star-forming galaxies \citep*[DSFGs;][]{Casey2014a}, one may expect a ubiquitous population of submillimeter-bright galaxies associated with protoclusters. 
Indeed, many protoclusters spanning $2<z<6$ have shown evidence for strong overdensities of massive galaxies with intensely high star formation rates, including the Spiderweb Galaxy protocluster at $z=2.16$ \citep{Dannerbauer2014a}, SSA22 at $z=3.09$ \citep{Steidel1998a, Chapman2009a, Umehata2015a}, the Distant Red Core  at $z=4.00$ \citep{Oteo2018a, Long2020a}, and SPT2349-56 at $z=4.3$ \citep{Miller2018a, Hill2020a}.
If protoclusters span tens of comoving Mpc at $z>2$ \citep{Casey2015a, Hung2016a, Chiang2015a}, is there a mechanism that could trigger simultaneous bursts of star formation across such wide scales \citep{Casey2016a}? 

However, with a small sample of DSFG-rich protoclusters, we cannot yet claim that they are ubiquitous -- indeed, mock catalogs derived from cosmological simulations have actually shown that submillimeter-selected galaxies are poor tracers of overdensities at $z\sim2$ \citep{Miller2015a}. 
There is instead the possibility that protoclusters  built their mass continuously at modest, sustained star formation rates, in which case one may expect to find more significant overdensities in star-forming galaxies traced by Ly$\alpha$ emitters (LAEs) and Lyman-break galaxies (LBGs). 
LAE overdensities have also been found within galaxy protoclusters at $z>2$ \citep[e.g.,][]{Steidel1998a, Chiang2015a, Toshikawa2018a, Jiang2018a}, and in some cases these trace the same structures as the more extreme DSFGs. 
Therefore, it is likely that galaxy protoclusters have complex galaxy populations that evolve along with the protoclusters themselves, so we must go beyond 
anecdotal identifiers of individual overdensities and begin to characterize the true effect of the dense environment on galaxy evolution. 

Here, we present new work on substructures associated with the larger ``Hyperion" protocluster at $z=2.47$ \citep{Cucciati2018a} which contains an exceptionally rare concentration of luminous AGN and DSFGs stretching across at least 50\,cMpc, with a prolific total SFR of $>$5000\,\sfr\, \citep{Diener2015a, Chiang2015a, Casey2015a}. 
\citet{Casey2015a} performed submillimeter stacking to derive the total mass of cold dust in the system and suggested that it was elevated above the field despite the individual galaxies displaying comparable SFRs. 
We obtained Very Large Array (VLA) observations of two dense knots along the transverse filaments in this structure to investigate whether these galaxies have correspondingly large molecular gas reservoirs. 
Characterizing the cold gas in protocluster member galaxies allows us to compare to field galaxies at similar redshift to see if the overdense environment affects star formation efficiency and molecular gas depletion. 

The paper is structured as follows: in Section \ref{sec:structure} we describe what is known about this structure at $z\sim2.5$; in Section \ref{sec:data} we present the VLA and ALMA observations in addition to available ancillary legacy data; in Section \ref{sec:anal} we present the images and spectra for the gas-rich sources in the filaments, followed by a discussion of the parent protocluster in Section \ref{sec:discussion}. 
We discuss the two filaments in Sections \ref{sec:pointing1} and \ref{sec:pointing2} and present our conclusions in Section \ref{sec:conclusions}.
We assume a concordance $\Lambda$CDM cosmology where $H_0 = 70$\,km\,s$^{-1}$\,Mpc$^{-1}$, $\Omega_{\Lambda} = 0.7$, and $\Omega_{m} = 0.3$.
We also adopt a Chabrier initial mass function \citep[IMF;][]{Chabrier2003a}.

\section{A Massive Star-forming Protocluster}
\label{sec:structure}

We report on two specific arcminute-scale sub-regions of a known $z=2.47$ protocluster structure located in the COSMOS field. 
The parent structure \citep[dubbed ``Hyperion" by][]{Cucciati2018a} was identified simultaneously in multiple surveys: as a $z=2.44$ overdensity of LAEs in the Hobby Eberly Telescope Dark Energy Experiment (HETDEX) Pilot Survey \citep{Chiang2015a}; as a few isolated 2--3\arcmin-scale overdensities at $z=2.45$ in the zCOSMOS survey using the Very Large Telescope \citep[VLT;][]{Diener2015a}; and as a large 20\arcmin-scale $z=2.47$ overdensity in a redshift survey of 450\um\, and 850\um-detected DSFGs in the \scubaii-imaged portion of the COSMOS field \citep{Casey2015a}. 
This overdensity was also reported in the Ly$\alpha$ tomography map of the COSMOS field \citep[CLAMATO;][]{Lee2016a, Lee2018a}.

\begin{figure}
    \includegraphics[width=1.0\columnwidth]{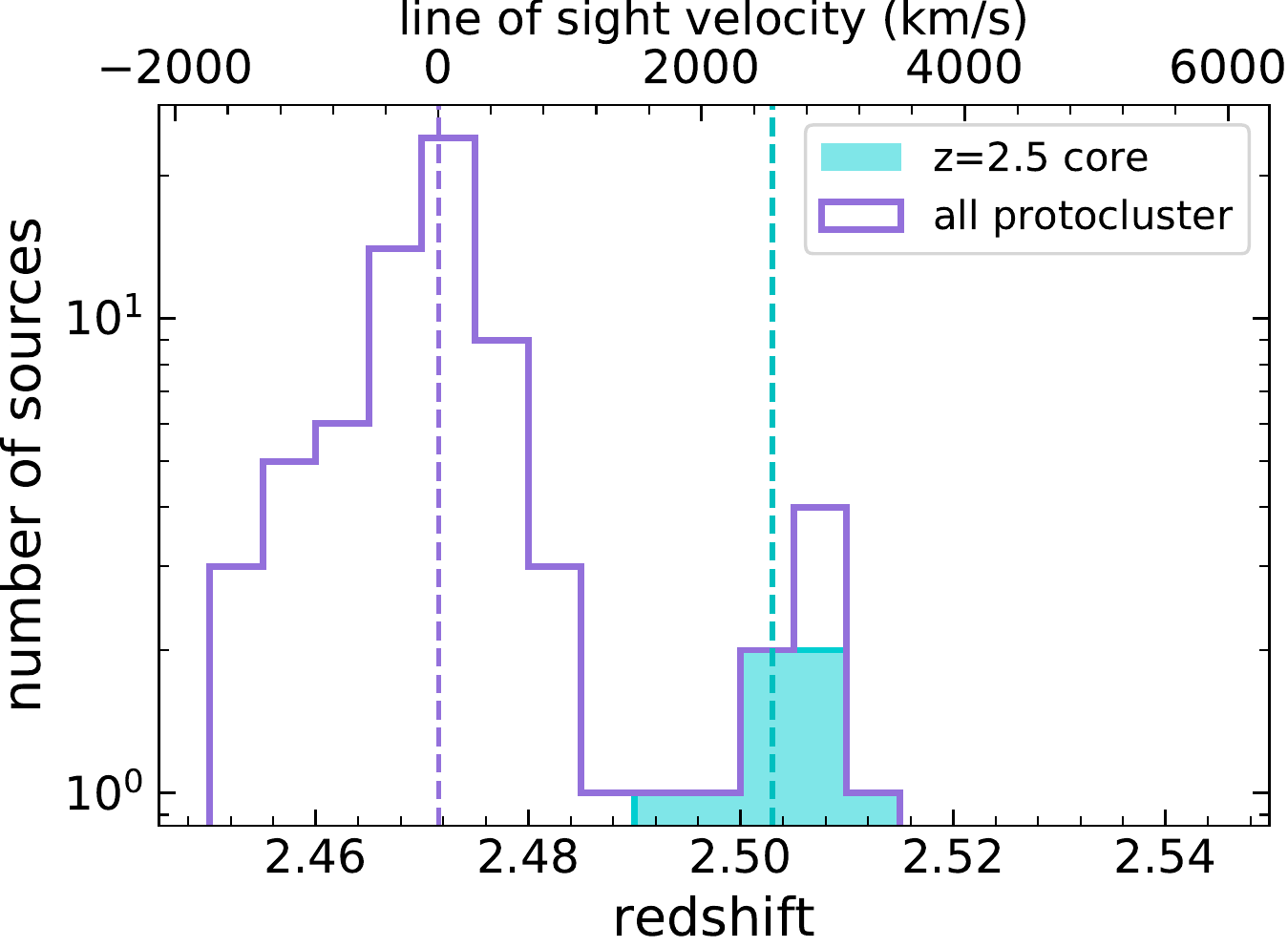}
    \caption{Purple histogram shows the redshift distribution for all spectroscopically confirmed sources in the $z=2.47$ Hyperion structure (purple dashed line); the filled cyan histogram shows the CO(1-0) detections reported here, at a $z=2.5$ overdensity within the structure (cyan dashed line). The nested core sits roughly 2200 km/s away from the center of the $z=2.47$ protocluster.}
    \label{fig:zhist}
\end{figure}

\citet{Zavala2019a} characterized the interstellar medium and stellar masses for 46 spectroscopically confirmed member galaxies using ALMA Band 6 continuum data and ancillary COSMOS data, finding that member galaxies have ISM and stellar masses $>10^{10}$\,\msun. 
The full structure is expected to exceed the mass of the Coma supercluster by $z=0$, crudely estimated to be (2 $\pm$ 1) $\times\, 10^{15}$\,\msun\, \citep{Casey2015a}. 
Given that new protocluster members are still being identified using various techniques, these estimates are considered lower limits. 
Figure \ref{fig:zhist} shows the redshift distribution of all currently known spectroscopically confirmed members of the $z=2.47$ protocluster, including the VLA detections presented here. 

In this paper we focus on the overdensity identified by \citet{Casey2015a} as a filamentary structure at $z=2.47$, called PCL1002, which contained seven DSFGs and five active galactic nuclei (AGN) in a comoving volume of 15,000\,Mpc$^3$ (an effective sky area of 200\,arcmin$^2$).  
Six DSFGs were spectroscopically confirmed as members of the structure through H$\alpha$ detections \citep{Casey2015a} and one via CO(1-0) \citep{Lentati2014a}.
Using a friends-of-friends algorithm, \citet{Casey2015a} additionally combined these with 35 UV-selected, spectroscopically-confirmed sources in the $z$COSMOS survey \citep{Lilly2009a} within $2.463<z<2.487$. 
PCL1002 is thus marked by overdensities of $\delta_{\rm DSFG} \approx 18$ and $\delta_{\rm LBG} \approx 3.3$, inferred by comparison to zCOSMOS spectroscopic yields at similar redshifts. 

Followup observations have revealed a particularly dense knot of dusty protocluster galaxies at $z=2.5$ which is likely associated with PCL1002/Hyperion \citep{Wang2016a, Wang2018a, Gomez-Guijarro2019a}. 
Depending on the selection criteria, between seven and fourteen CO detections have been reported within $\sim 20\arcsec$ and $\Delta z = 0.02$, with molecular gas reservoirs of $\rm M_{H_2} > 10^{10}$\,$\rm M_{\odot}$. 
With a spatially coincident X-ray detection extending roughly 20\arcsec, \citet{Wang2016a} cites this structure as a virialized galaxy cluster with an established hot ICM. 
In this paper we present deeper VLA observations of this core as well as new VLA data centered on another filament of the $z=2.47$ structure approximately 6 arcmin away, and revisit the physical characterization and interpretation of the radio and X-ray sources in this structure.
We refer the reader to Section \ref{sec:discussion} and Figure \ref{fig:3d} for a discussion of the $z=2.5$ core in context with the larger parent structure of Hyperion.

\section{Data Reduction and Imaging}\label{sec:data}

\subsection{VLA Data}\label{sec:VLA}
The data were taken with the Karl G. Jansky Very Large Array in D configuration in the Ka band (33\,GHz). 
The observations as part of VLA program 15B-210 (PI: C. Casey) were taken in eight epochs in 2015 November, with seven six-hour integrations and one four-hour integration, for a total of 46 hours.
Each of the two pointings was observed for 12.5 hours on-source.
We used the four 8-bit samplers (each covering a bandwidth of 1024\,MHz) with 15\,MHz binning in 128\,MHz sidebands. 
The sampler pairs, A0/C0 and B0/D0, were tuned to 33.25 and 33.20\,GHz respectively. 
Thus, the effective frequency coverage is 32.35$-$33.29\,GHz, corresponding to $2.46<z<2.56$ for the ground state transition of carbon monoxide, $^{12}$CO($J$=1-0). 

Pointing 1 is centered at $\alpha$\,=\,10:00:57.30 $\delta$\,=\,+02:20:13.3, selected to include a \scubaii-selected galaxy 
named 450.28 \citep{Casey2013a}. 
Prior to high-resolution ALMA follow-up, 450.28 was identified at a redshift of $z=2.47$ via \halpha\, emission \citep{Casey2015a}; we now know that this spectroscopic redshift is only one of multiple sources at similar redshifts.
450.28 is also associated with an extremely bright (S$_{350\mu m} \sim$\,90\,mJy), unresolved \herschel/SPIRE counterpart. 
We combine the total VLA time on source for this pointing (12.5\,hr) with four sets of observations from VLA program 15B-290 (PI: T. Wang) for an effective integration time of 26\,hr. 
Pointing 2 is centered at  $\alpha$\,=\,10:00:34.6  $\delta$\,=\,+02:22:01.20 to include another \scubaii-selected galaxy from \citet{Casey2013a} named 450.58.
Both pointings were selected for their relative spatial overdensities of DSFGs and LBGs with reliable redshifts at $2.46<z<2.55$. 
We use the Common Astronomy Software Applications \citep[CASA;][]{McMullin2007a}\footnote{\url{https://casa.nrao.edu/index.shtml}} and the 2016 VLA pipeline to reduce the full dataset. 
The data cubes have a channel width of 70\,\kms\, 
and a pixel scale of $0\farcs42$ per pixel ($\approx3.5$\,kpc at $z=2.5$). 
The images are produced via the task \textsc{clean}. 
We use natural weighting to maximize the signal to noise of expected point sources, with a final synthesized beam size of $2\farcs87\times2\farcs56$. 
We trim the images to 40\% of the primary beam response (FWHP size $=1.35\arcmin$). 
We later find that the CO-identified sources are very close together with complex morphology, so we repeat the analysis in both cleaned and dirty maps to account for potential cleaning artifacts. 
The typical rms per channel is 10 $\mu$Jy/beam. 

\subsection{ALMA Data}\label{sec:alma}
\subsubsection{Band 6 data}
ALMA Band 6 continuum observations were conducted on 4 April 2017 as part of the Cycle 4 program 2016.1.00646.S (PI: C. Casey) aimed at observing the millimeter continuum across the full $z=2.47$ structure. 
Full details of the observations and data reduction can be found in \citet{Zavala2019a}, but we briefly summarize them here. 

The data were taken using the 12\,m antennae in a compact configuration (maximum baseline of 0.46\,km),  
with an average on-source integration time of 5\,min per source. 
These observations were targeted towards detecting the ISM continuum in previously identified DSFGs, so the correlators were configured in order to maximize the continuum sensitivity. 
Centered at 1.2\,mm, the observations have a total bandwidth of 7.5\,GHz with 31\,MHz channels.  
The observations were reduced in CASA using the ALMA reduction pipeline; natural weighting and a pixel size of 0$\farcs$12 were used to obtain a synthesized beam-size of $\theta_{\rm FWHM}\approx0\farcs9$ and an average spectral noise of $45\,\mu$Jy/beam. 
Of our two VLA pointings, only the 450.28 field (Pointing 1) has Band 6 continuum coverage, so we do not have dust measurements for the 450.58 field (hereafter Pointing 2).

\subsubsection{Band 3 data}
We additionally use ALMA Band 3 observations from Cycle 3 (program code 2015.1.00207.S, PI: C. Casey) intended to detect CO(3-2) in six protocluster DSFGs. 
These data were taken using the 12\,m antennae in a compact configuration, with an average on-source integration time of 25\,min and a minimum baseline of 64\,m. 
Two of the pointings provide coverage of each of the VLA fields presented here, but with only one available spectral window which covers 98.7-100.5\,GHz ($2.44<z<2.50$) in 31\,MHz bins. 
However, the center of the protocluster core discussed herein has now been updated from $z=2.47$ to $z=2.5$, so most of the galaxies in VLA Pointing 1 are either not covered or fall on the edge of the spectral window. 
Therefore, we only report CO(3-2) upper limits for sources falling inside the covered bandwidth.

These data were reduced in CASA using the ALMA reduction pipeline. 
The images are cleaned using natural weighting and a uvtaper of 2\farcs5 (to maximize sensitivity to possible spatially extended dust emission), resulting in a synthesized beam size of 0\farcs87$\times$0\farcs86. 
Since the sources are expected to be faint we do not perform continuum subtraction. 
We use a pixel size of 0\farcs15 and a channel width of 70\,\kms\, to match the spectral resolution of the VLA data.
The average rms/channel is 0.9 mJy/beam. 

\subsection{Ancillary COSMOS data}\label{sec:cosmos}

Since this protocluster lies in the COSMOS field, the CO and dust continuum observations presented here are complemented by a wealth of multi-wavelength data taken as part of the COSMOS survey \citep{Scoville2007a, Capak2007a}. 
Where the sources can be resolved individually, we match CO-identified sources to the COSMOS 2015 catalog \citep{Laigle2016a} with a match radius of 1$\farcs$5 for more complete optical/near-infrared (OIR) constraints.

In Section \ref{sec:anal}, we model the individual OIR SEDs using \magphys\, \citep{daCunha2008a, daCunha2015a} using the following photometry:  Subaru Suprime-Cam broad-band and 
narrow-band 
filters; UltraVISTA \textit{YJHKs}; $Spitzer$/IRAC 3.5, 4.5, 5.8, and 8.0\,\um; and $Spitzer$/MIPS 24\,\um. 

We also search for individual counterparts in the two major radio continuum surveys included in COSMOS: the VLA-COSMOS 1.4\,GHz Deep project \citep[with a spatial resolution of 1\farcs5 and depth 7\,$\mu$Jy/beam;][]{Schinnerer2007a},  and the 3\,GHz project \citep[angular resolution 0\farcs7 and depth 2.3\,$\mu$Jy/beam;][]{Smolcic2017a}. 
These radio counterparts are meant to account for possible contribution to the infrared flux from radio AGN as well as star formation-generated radio emission.

Additionally, there are \herschel\, PACS (100\,\um\, and 160\,\um) and SPIRE (250\,\um, 350\,\um, and 500\,\um) data available which we use to model the FIR SED of the protocluster core. 
However, the individual protocluster sources are confused by SPIRE's large beam (36\arcsec\, at 500\,\um), so we fit the FIR SED treating the core as a single source and calculate associated physical properties in aggregate.

Finally, we check the \textit{Chandra} point source catalog from \citet{Civano2016a} for potential X-ray AGN. 
We also check for extended X-ray emission using the merged COSMOS \textit{XMM} + \textit{Chandra} maps for consistency with the analysis of \citet{Wang2016a}. 

\section{Data Analysis} \label{sec:anal}

\subsection{CO(1-0) Source Extraction}\label{co10}
We search for CO(1-0) sources using the positions of previously identified sources in COSMOS as a prior. 
Where available, we use the \halpha-detected sources from \citet{Casey2015a} and Band 6 continuum sources \citep{Zavala2019a} as positional priors, for a total of twelve priors. 
We searched for additional sources (i.e., not based on these priors) in the dirty cube by iteratively creating moment-0 maps integrated along five consecutive channels at a time, corresponding to an effective line widths of 350\,km\,s$^{-1}$ \citep[typical for CO lines at high redshift;][]{Bothwell2013a}.
In some cases, we find that the CO detections overlap spatially when integrated, so we use their offsets in frequency to distinguish between sources, and we count the number of detections by the number of discrete bright peak pixels.

In total, we report seven CO(1-0) detections in Pointing 1 and one in Pointing 2 (Table \ref{table:co}).
Note that all CO sources reported in Pointing 1 have also been confirmed at shallower depth in other publications \citep{Wang2016a, Wang2018a, Gomez-Guijarro2019a}, though each study uses different SNR and size thresholds so the number of independent sources may vary.
At this spatial resolution, the sources are mostly unresolved, but some flux is missed in a beam-sized aperture.
We therefore extract spectra using a manually-placed aperture that encompasses flux from adjacent pixels which all have SNR$>$3 (typically the size of 1--2 VLA beams), taking care to shape the apertures such that we do not double-count flux from nearby sources.
In Appendix \ref{app:extraction} we describe in more detail our source extraction method and test whether the sources are spatially resolved.

\begin{figure*}
\begin{center}
 \gridline{\fig{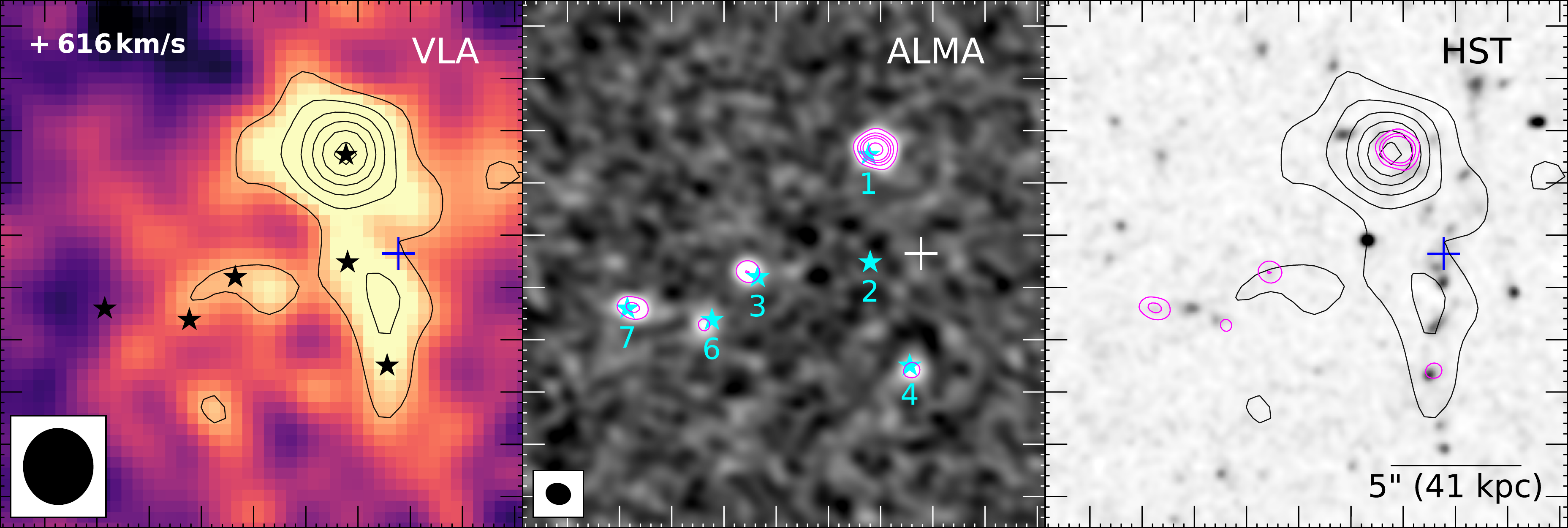}{1.8\columnwidth}{}}
 \vspace{-1cm}
 \gridline{\fig{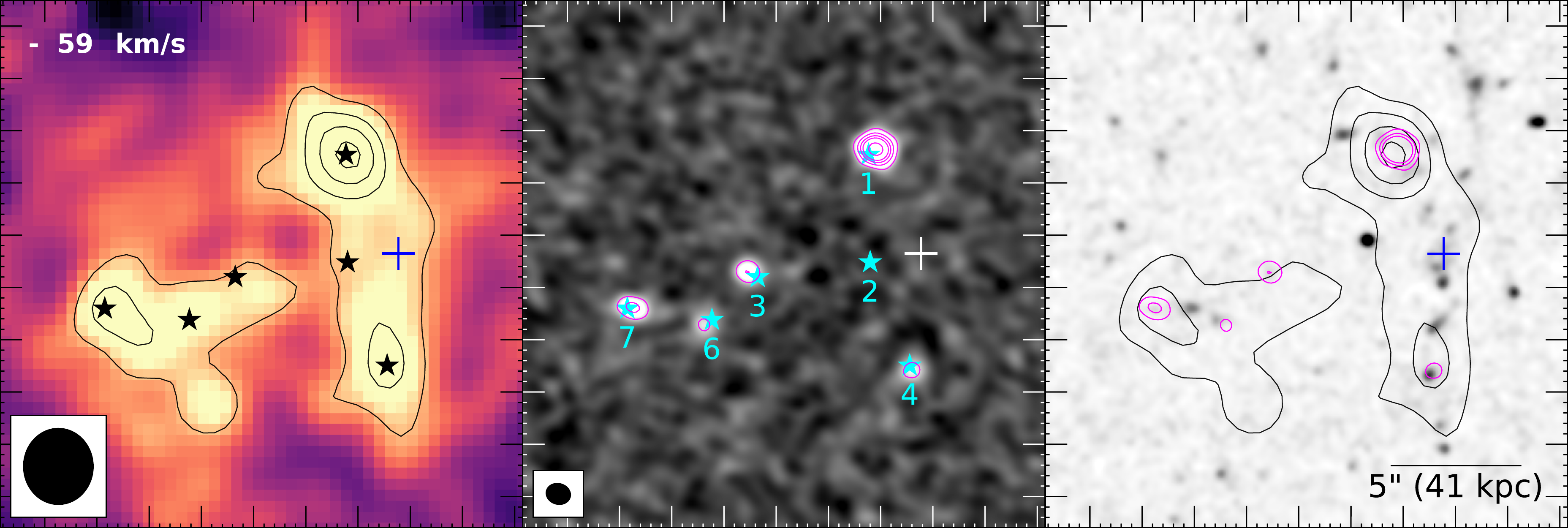}{1.8\columnwidth}{}}
 \vspace{-1cm}
 \gridline{\fig{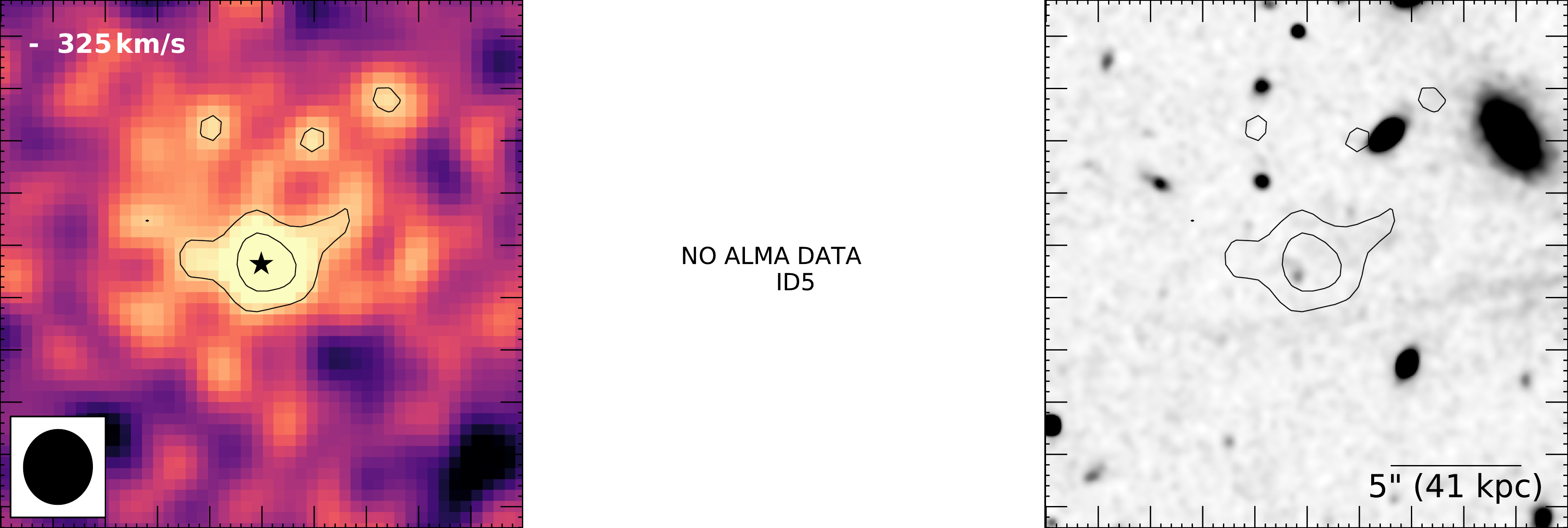}{1.8\columnwidth}{}}
 \vspace{-1cm}
 \gridline{\fig{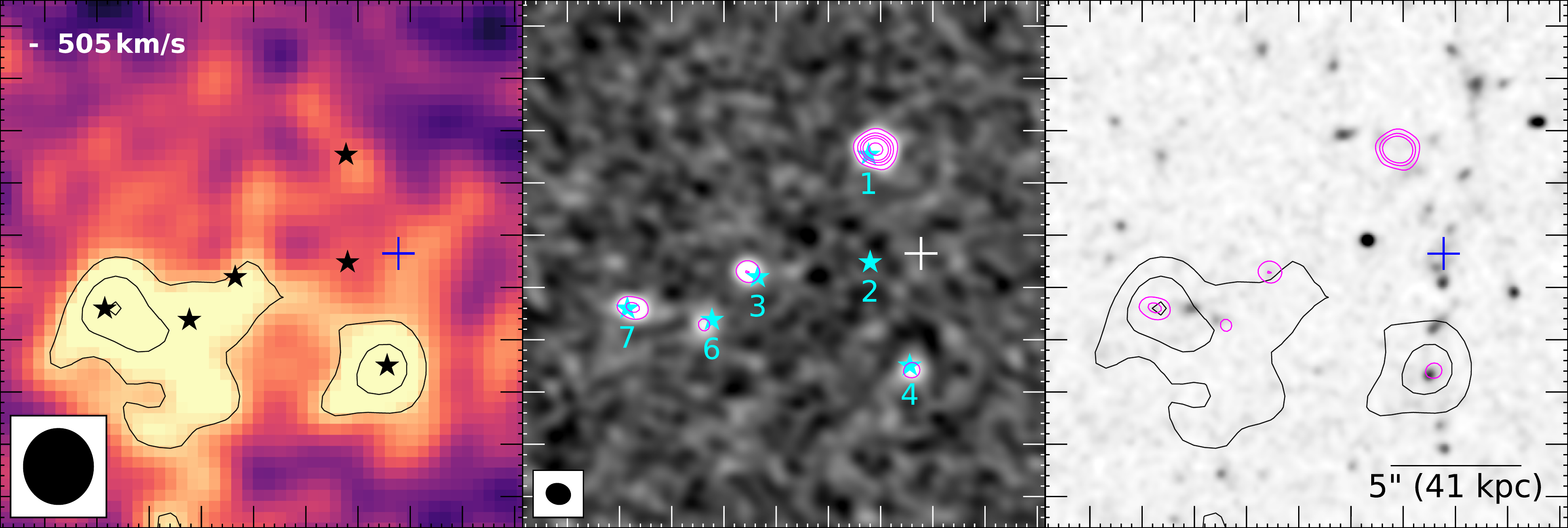}{1.8\columnwidth}{}}
 \vspace{-1cm}
 
\end{center}
\caption{20\arcsec$\times$20\arcsec\, maps of all seven sources in CO, dust continuum, and IR continuum at four representative velocity slices with respect to $z=2.503$.
In all panels, the large cross symbol indicates the center of the VLA pointing, and the stars indicate the centroids of the CO sources, with the corresponding ID numbers in panel 2 (note that ID5 is approximately 45\arcsec\, from the pointing center).
\textit{Left:} Integrated moment-0 maps in Pointing 1, with the velocity offset indicated in the upper left. 
Contours begin at 3$\sigma$ and increase in steps of 1$\sigma$, and the VLA beam is displayed in the lower left. 
Center: ALMA Band 6 continuum signal-to-noise maps centered on the same position as left panel, with ALMA beam displayed in the lower left. 
Magenta contours begin at 4$\sigma$ and increase in steps of 4$\sigma$ until 12$\sigma$, then in steps of 10$\sigma$. 
\textit{Right:} \textit{HST} F814W images, with VLA moment-0 contours in black and ALMA contours in magenta. }
\label{fig:m0}
\end{figure*}

\begin{figure*}
\begin{center}
    \includegraphics[width=2.0\columnwidth]{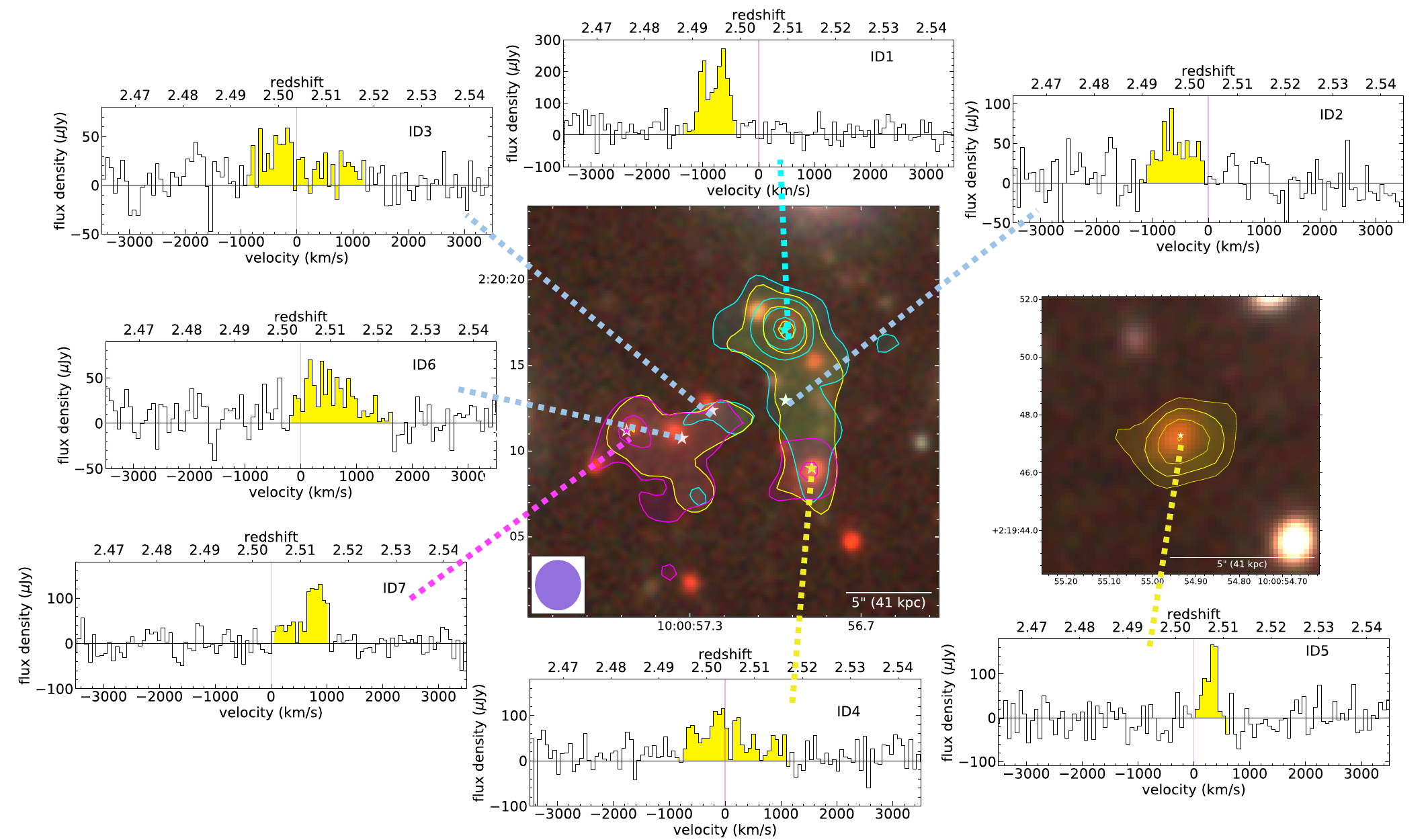}
    \caption{
    Zoomed-out VLA moment 0 contours integrated in three representative redshift slices, overlaid on NIR counterparts.
    The dashed lines connect the individual spectra to the locations of the 7 CO centroids.
    The colors correspond to the three redshift slices (cyan is $z=2.49$, yellow is $z=2.50$, magenta is $z=2.51$). 
    The two background images are composite UltraVISTA images in $Y$, $J$, and $K_s$ bands. 
    At left is a 24\arcsec\, cutout encompassing IDs 1, 2, 3, 4, 6 and 7, and at right is a 10\arcsec\, cutout of ID5, about 43\arcsec\, away from the other detections.}
    \label{fig:uvista}
\end{center}
\end{figure*}

In Figure \ref{fig:m0}, we show zoomed-in moment-0 maps from the VLA at four velocity slices with respect to a central redshift of $z=2.503$, plus ALMA continuum and \textit{HST} maps. 
While the dust emission in each source is relatively compact, the marginally extended nature of very nearby CO(1-0) sources combined with a heterogeneous set of line profiles points to a complicated overall molecular gas morphology. 
The apparent extended nature of the molecular gas along the frequency/velocity axis could imply either that there is significant gas interaction between galaxies, or that we are tracing individual peculiar velocities in a single halo; we discuss this further in Section \ref{sec:discussion}.
Figure \ref{fig:uvista} shows the full zoomed-out moment-0 maps overlaid on NIR images from UltraVISTA at the same velocity slices, along with their corresponding spectra, where one can clearly see the ambiguous and overlapping nature of the molecular gas reservoirs. 

\begin{figure}
\includegraphics[width=1.0\columnwidth]{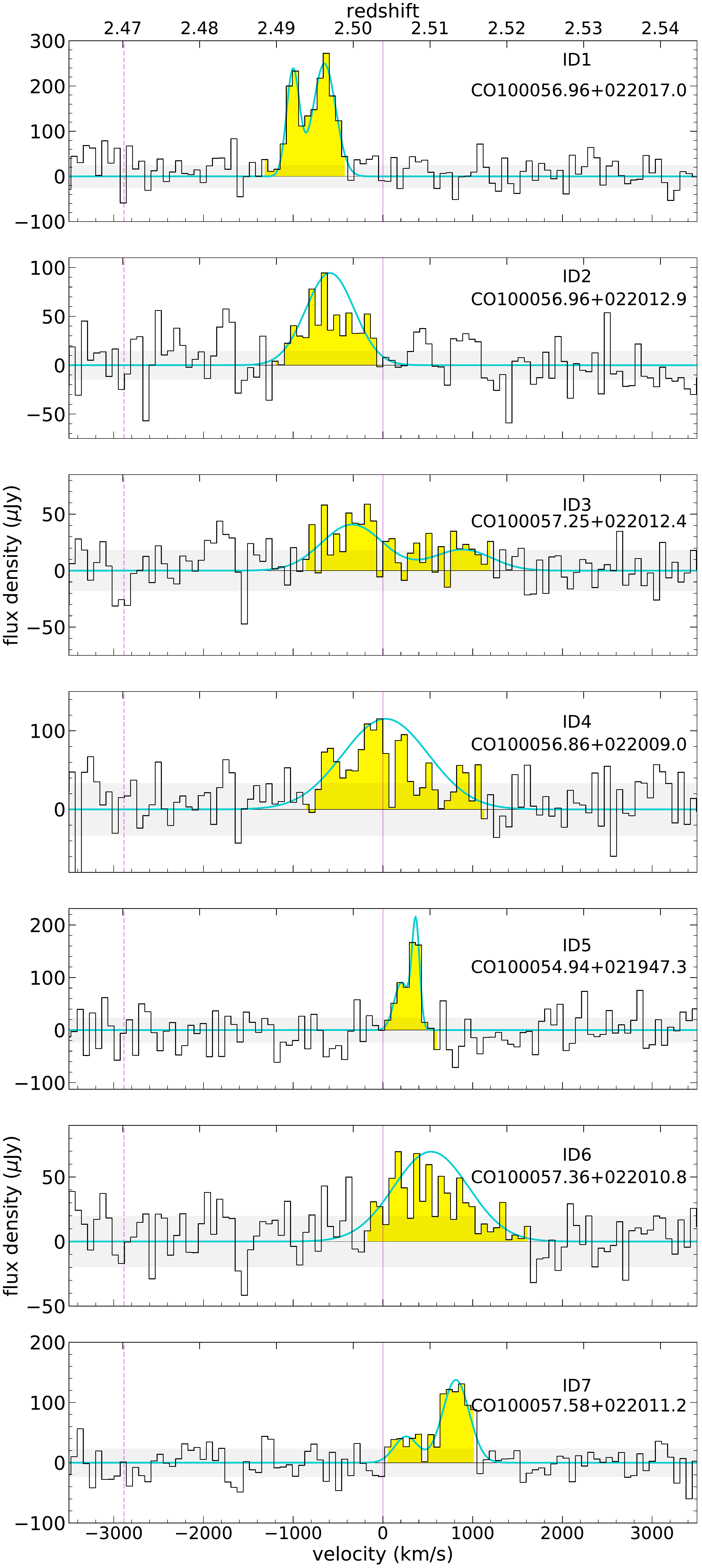}
\caption{CO(1-0) spectra for the seven detected sources in Pointing 1. 
The channels along which the flux is integrated are highlighted in yellow. 
The equivalent Gaussian curve is overlaid in cyan (see text for details). 
The rms per channel is shown in shaded grey. 
The solid magenta line denotes $z=2.503$, the central redshift of the nested core structure, while the dotted magenta line indicates $z=2.47$, the center of the larger protocluster.}
\label{fig:spectra}
\end{figure}

We then calculate the spectral line moments to derive molecular gas masses from the spectra shown in Figure \ref{fig:spectra}. 
None of the spectra show evidence of continuum emission, so we do not perform any continuum subtraction. 
Some CO lines appear to be comprised of both broad and narrow components or double-peaked lines.
Therefore, we directly calculate the spectral line moments, allowing us to constrain the redshift, FWHM, and total line intensity without having to assume each CO line is comprised of a single Gaussian component.
In Appendix \ref{app:moments} we outline these calculations in detail, roughly following the methods of \citet{Bothwell2013a}.

\movetabledown=2.5in
\begin{rotatetable}
\begin{deluxetable*}{cccccccccccccc}
\tablecaption{Details of CO data and derived gas properties. \label{table:co}}
\tablehead{\colhead{Name$^1$} & \colhead{RA} & \colhead{Dec} & \colhead{$z$} & \colhead{I$_{\rm CO(1-0)}$} & \colhead{FWHM} & \colhead{SNR$^2$} & \colhead{log(L$^{\prime}_{\rm CO(1-0)}$)} & \colhead{M$_{\rm gas}^3$} & \colhead{I$_{\rm CO(3-2)}$} & \colhead{FWHM} & \colhead{SNR} & \colhead{log(L$^{\prime}_{\rm CO(3-2)}$)} & \colhead{r$_{31}^4$} \\
\colhead{} & \colhead{(J2000)} & \colhead{(J2000)} & \colhead{} & \colhead{(mJy\,km/s)} & \colhead{(km/s)} & \colhead{} & \colhead{(K\,km/s\,pc$^2$)} & \colhead{(10$^{10}$ M$_{\odot}$)} & \colhead{(mJy\,km/s)} & \colhead{(km/s)} & \colhead{} & \colhead{(K \kms pc$^2$)} & \colhead{}}
\startdata
ID1 & 150.237350 & 2.338078 & 2.494 & 122.0 $\pm$ 6.4 & 486 $\pm$ 36 & 16.04 & 10.55 $\pm$ 0.02 & 22.96$_{-0.74}^{+0.74}$ & 1261.0 $\pm$ 143.0 & 532 $\pm$ 78 & 11.26 & 10.61 $\pm$ 0.05 & 1.14 $\pm$ 0.14 \\
ID2 & 150.237333 & 2.336935 & 2.496 & 39.0 $\pm$ 3.9 & 532 $\pm$ 54 & 4.92 & 10.05 $\pm$ 0.05 & 7.35$_{-0.45}^{+0.45}$ & $<$ 190 & $<$ 500 & -- & $<$ 9.79 & $<$ 0.54 \\
ID3 & 150.238529 & 2.336777 & 2.503 & 46.3 $\pm$ 6.9 & 1271 $\pm$ 143 & 3.96 & 10.13 $\pm$ 0.07 & 8.76$_{-0.81}^{+0.81}$ & -- & -- & -- & -- & --  \\ 
ID4 & 150.236912 & 2.335836 & 2.504 & 92.6 $\pm$ 12.6 & 1095 $\pm$ 135 & 7.74 & 10.43 $\pm$ 0.06 & 17.53$_{-1.47}^{+1.47}$ & -- & -- & -- & -- & --  \\ 
ID5$^5$ & 150.228896 & 2.329801 & 2.507 & 37.4 $\pm$ 5.1 & 172 $\pm$ 71 & 9.97 & 10.04 $\pm$ 0.06 & 7.10$_{-0.59}^{+0.59}$ & -- & -- & -- & -- & -- 
\\
ID6 & 150.239017 & 2.336322 & 2.509 & 50.0 $\pm$ 7.0 & 981 $\pm$ 145 & 4.81 & 10.16 $\pm$ 0.07 & 9.50$_{-0.82}^{+0.82}$ & -- & -- & -- & -- & -- \\ 
ID7 & 150.239917 & 2.336443 & 2.511 & 64.1 $\pm$ 6.4 & 619 $\pm$ 56 & 8.62 & 10.27 $\pm$ 0.05 & 12.19$_{-0.75}^{+0.75}$ & -- & -- & -- & -- & -- \\ 
\hline
450.58 & 150.150250 & 2.364176 & 2.464 & 180.5 $\pm$ 19.1 & 334 $\pm$ 59 & 9.87 & 10.70 $\pm$ 0.03 & 32.94 $\pm$ 2.64 & $<$ 1400 & $<$ 500 & $<$ 2 & $<$ 10.65 & $<$ 0.90
\enddata
\tablecomments{1. Names beginning with ``ID" are labelled in order of increasing redshift in Pointing 1; 450.58 is in Pointing 2. 2. SNR is the ratio of integrated moment-0 flux density to the rms in the moment-0 maps. 3. M$_{\rm gas}$ is calculated assuming $\alpha_{\rm CO}$ = 6.5\acounits, as advocated for in Appendix \ref{app:aco} rather than the typical value of $\alpha_{\rm CO} = 1$ assumed for high-redshift DSFGs \citep*[e.g.,][]{Carilli2013b}. 4. Ratio of L$^{\prime}_{\rm CO(3-2)}$/L$^{\prime}_{\rm CO(1-0)}$. 5. Identified as 850.82 in \citet{Casey2013a}.}
\end{deluxetable*}
\end{rotatetable}
\clearpage

After measuring the line intensity $I_{CO}$, the line luminosity $L^{\prime}_{\rm CO}$ is the following:

\begin{equation}
    L^{\prime}_{CO} = 3.25\times10^7\times S_{\nu}\Delta v \frac{D_L^2}{(1+z)^3 \nu_{obs}^2} \rm K\,km\,s^{-1}\,pc^2
\end{equation}

\noindent where $S_{\nu}\Delta v = I_{CO}$ is measured in Jy \kms, the luminosity distance $D_L$ in Mpc, and $\nu$ in GHz.

Finally, we calculate the molecular gas mass, $M_{\rm gas} = \alpha_{CO} L'_{CO}$, where \alphaco\,=6.5\,\acounits\, is the empirical factor to convert to gas mass from line luminosity.
In Appendix \ref{app:aco} we motivate this choice of \alphaco\,, which is higher than the value of \alphaco$\approx$1\,\acounits\, that has historically been assumed for high-redshift galaxies.

\subsection{CO(3-2) Source Extraction}\label{sec:co32}
\begin{figure}
\begin{center}
\includegraphics[width=1.0\columnwidth]{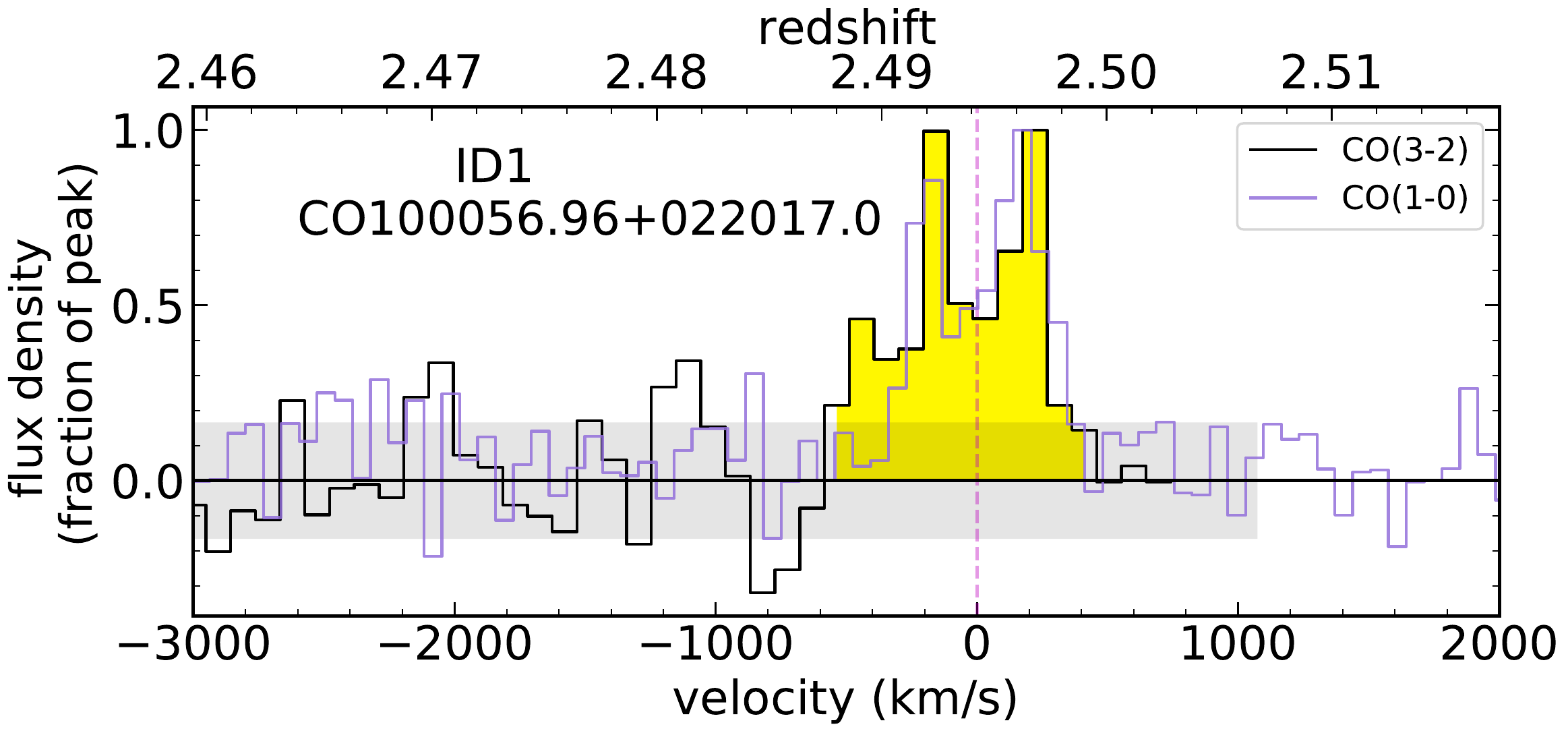}
\includegraphics[scale=0.5]{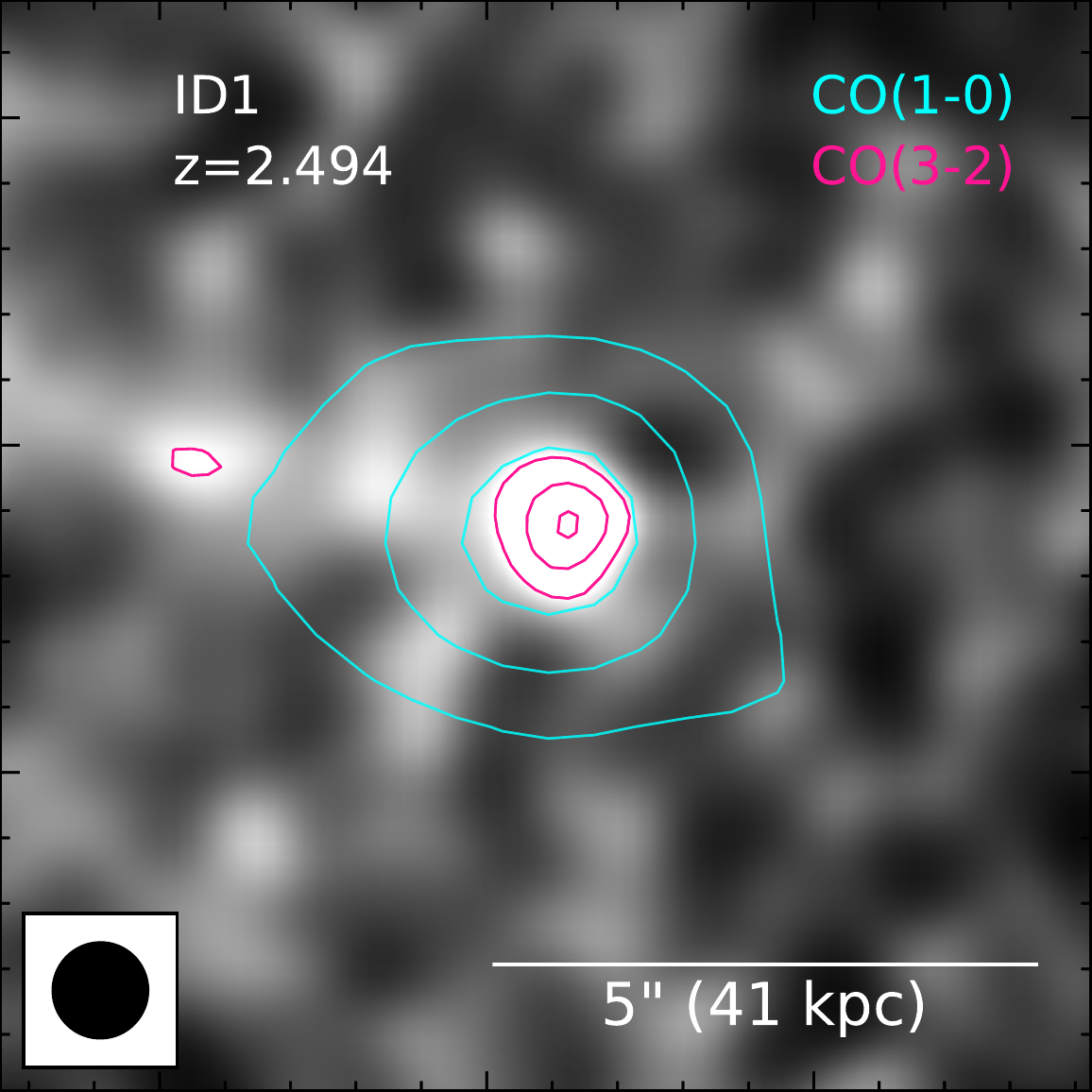}
\caption{Analysis for the only CO(3-2) detection in the core, corresponding to ID1.
\textit{Top:} ALMA Band 3 spectrum in black, scaled to the peak flux and with velocity offset corresponding to the center of the CO(1-0) spectrum (purple). 
The rms/channel of the ALMA data is shaded in grey --- note that this detection is on the edge of the spectral window. 
\textit{Bottom:} Smoothed 10\arcsec ALMA moment-0 map, with magenta contours (3--9$\sigma$) corresponding to the CO(3-2) detection, and cyan contours (5, 10, 15$\sigma$) corresponding to CO(1-0).}
\label{fig:co32}
\end{center}
\end{figure}

We repeat the search for line detections in the ALMA Band 3 data following the procedures outlined in Section \ref{co10}. 
Since the observations were tuned to a central redshift of $z=2.47$, only two galaxies in VLA Pointing 1 and one in Pointing 2 have the frequency coverage to detect CO(3-2). 
Of these, only ID1 is detected in CO(3-2), with SNR of 9. 
We report the detections and upper limits in Table \ref{table:co}. 
We find that the ratio of (3-2) to (1-0) line luminosities, $r_{31}$, = $1.37 \pm 0.20$ for ID1 and an upper limit of $r_{31} < 0.72$ for ID2, suggesting a thermalized state of the ISM in line with $r_{31} = 0.90 \pm 0.40$ found for combined populations of submillimeter star-forming and AGN host galaxies at $2<z<3$ \citep{Sharon2016a, Kirkpatrick2019a}. 
The CO(3-2) emission in ID1 appears slightly resolved by ALMA and more compact than the CO(1-0) emission (see Figure \ref{fig:co32}), although since the CO(1-0) sources are unresolved, higher resolution observations are needed to verify this. 
Slightly more compact emission in CO(3-2) is expected since the more highly-excited gas is likely isolated to small molecular clouds throughout the galaxy, but it is more likely due to the poor resolution of the VLA, since AlMA's beamsize ($\theta\approx0$\farcs9) corresponds to roughly 8 kpc at this redshift.

\subsection{Dust Masses}\label{dustmass}
We use the ALMA Band 6 map centered on Pointing 1 to derive millimeter continuum flux densities. 
Using the CO positions as priors, we search the non-primary beam-corrected maps for sources with SNR\, \gt\, 3. 
ID5 lies outside of the Band 6 primary beam. Of the remaining six sources, five are detected at 1.2\,mm. 
Dust continuum detections at this wavelength correspond to $\lambda \approx$ 300\,\um\, which is on the Rayleigh-Jeans tail of the blackbody emission, where the dust emission is assumed to be optically thin. 
We can then calculate dust masses directly since it is proportional to the observed millimeter flux and the dust temperature, which we do using the framework described for 850\,\um sources in \citet{Dunne2000a}:

\begin{equation}
   \rm  M_{dust} = \frac{S_{\nu_{obs}} D_L^2 (1+z)^{-(3+\beta)}}{\kappa(\nu_{ref}) B_{\nu}(\nu_{ref}, T_{dust})} \Big{(}\frac{\nu_{ref}}{\nu_{obs}}\Big{)}^{2+\beta} \Big{(}\frac{\Gamma_{RJ(ref,0)}}{\Gamma_{RJ}}\Big{)}
\end{equation}

The emissivity, $\beta$, is fixed at 1.8. 
The quantity $\kappa_{\nu}$ is the dust mass absorption coefficient evaluated at rest-frame $\nu_{\rm ref}$; here we use $\kappa$(450\,\um) = 1.3$\pm$0.2\,cm$^2$\,g$^{-1}$ \citep{Li2001a} which is 1.5\,mm in the observed frame. 
$B_{\nu}$ is the Planck function evaluated at mass-weighted dust temperature $T_d$ which we fix at 25\,K as in \citet{Scoville2016a}. 
Finally, $\Gamma_{\rm RJ}$ is the Rayleigh-Jeans correction factor. 
The quantities derived from these fluxes are listed in Table \ref{table:alma} and the maps are displayed in the center panels of Figure \ref{fig:m0}.

\begin{deluxetable}{cccc}
\tablehead{\colhead{Name} & \colhead{S$_{\rm 1.2mm}$} & \colhead{M$_{\rm dust}$} & \colhead{G/D ratio} \\
\colhead{} & \colhead{(mJ/beam)} & \colhead{(10$^8$\,M$_{\odot}$)} & \colhead{}}
\tablecaption{ALMA Band 6 derived properties.\label{table:alma}}
\startdata
ID1 & 2.27 $\pm$ 0.06 & 27.0 $\pm$ 4.2 & 85 $\pm$ 13 \\
ID2 & $<$0.125 & $<$1.5 & $>$300 \\
ID3 & 0.47 $\pm$ 0.06 & 5.6 $\pm$ 1.2 & 156 $\pm$ 32 \\
ID4 & 0.34 $\pm$ 0.06 & 4.0 $\pm$ 0.9 & 436 $\pm$ 102 \\
ID5 & -- & -- & -- \\
ID6 & 0.29 $\pm$ 0.07 & 3.4 $\pm$ 1.0 & 280 $\pm$ 85 \\
ID7 & 0.95 $\pm$ 0.09 & 11.3 $\pm$ 2.0 & 108 $\pm$ 19
\enddata
\tablecomments{The dust masses have been calculated directly from S$_{\rm 1.2 mm}$ (see text for details). 
Note that ALMA Band 6 data has not been taken for the 450.58 pointing, and ID5 falls out of the Band 6 primary beam for Pointing 1.}
\end{deluxetable}

\begin{deluxetable}{cccc}
\tablecaption{Near-IR photometry and derived stellar masses. \label{table:phot}}
\tablehead{\colhead{Name} & \colhead{$K_s$} & \colhead{$H$} & \colhead{M$_{\star}$}\\
\colhead{} & \colhead{($\mu$Jy)} & \colhead{($\mu$Jy)} & \colhead{(10$^{11}$ M$_{\odot}$)}}
\startdata
ID1 & 4.98 $\pm$ 0.18 & 7.06 $\pm$ 0.20 & 2.1$_{-1.5}^{+2.8}$ \\
ID2 & 1.85 $\pm$ 0.18 & 2.10 $\pm$ 0.20 & 0.1$_{-0.0}^{+0.1}$ \\
ID3 & 2.59 $\pm$ 0.19 & 3.99 $\pm$ 0.21 & 1.0$_{-0.6}^{+0.8}$ \\
ID4 & 6.69 $\pm$ 0.18 & 10.89 $\pm$ 0.20 & 2.9$_{-1.7}^{+1.7}$ \\
ID5 & 2.54 $\pm$ 0.17 & 4.89 $\pm$ 0.17 & 1.9$_{-1.1}^{+0.8}$ \\
ID6 & 3.43 $\pm$ 0.19 & 6.44 $\pm$ 0.21 & 1.6$_{-0.8}^{+1.3}$ \\
ID7 & 1.97 $\pm$ 0.18 & 3.22 $\pm$ 0.20 & 1.6$_{-0.9}^{+0.9}$ \\
450.58 & 4.00 $\pm$ 0.17 & 7.10 $\pm$ 0.16 & 1.7$_{-1.1}^{+1.8}$
\enddata
\end{deluxetable}

\begin{figure}
 \includegraphics[width=1.0\columnwidth]{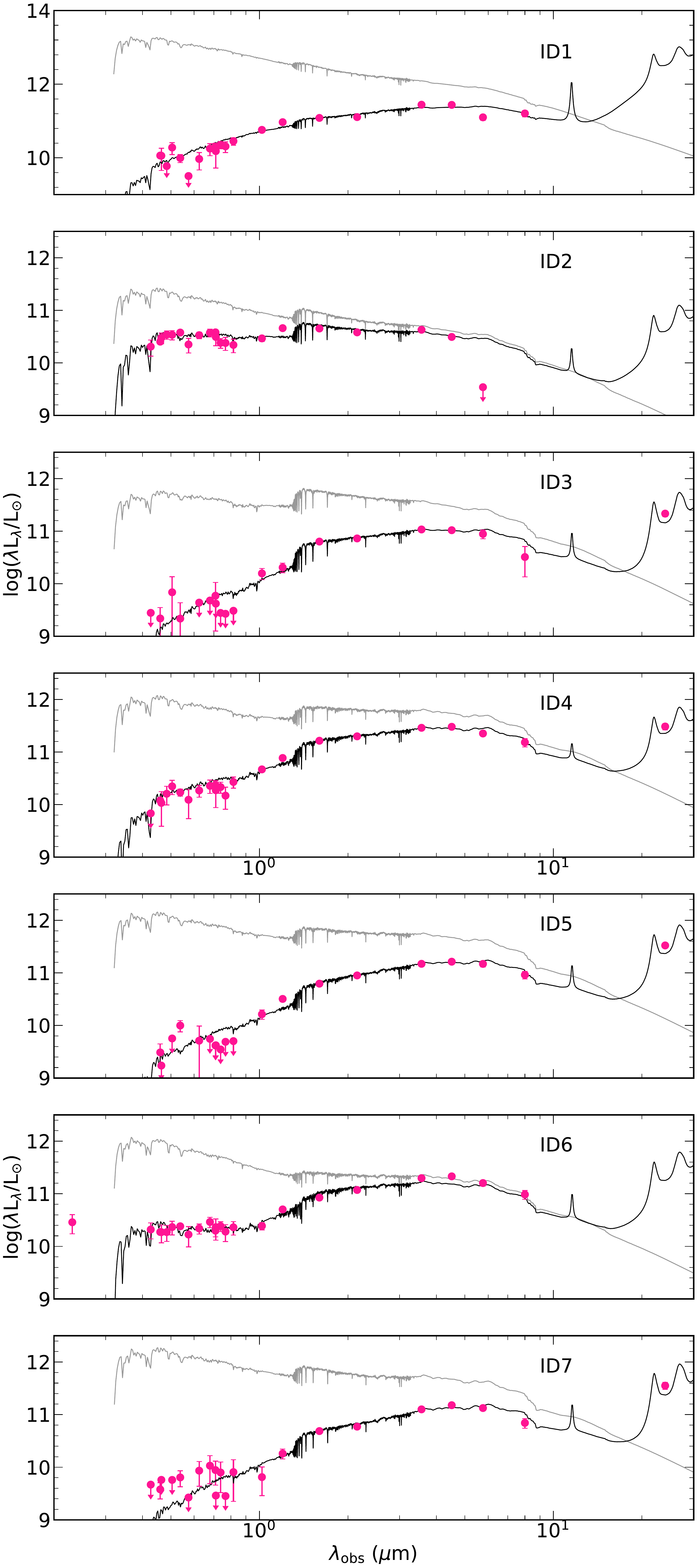}
 
 \caption{Spectral energy distributions of the protocluster core galaxies fit using High-Z \textsc{magphys}. 
 The pink data points correspond to COSMOS photometry \citep{Laigle2016a}. 
 The grey curve is the unattenuated spectrum, while the black curve is the observed (attenuated) SED. 
 These are fit using data blueward of 24\,\um, as the sources are spatially confused at longer wavelengths by \textit{Herschel/SPIRE}, so we calculate an aggregate FIR SED separately (see text for details). 
 Note that ID2 is not detected at 5.7, 7.8, or 24\,\um, and \textsc{magphys} does not use upper limits.
    Downward arrows indicate upper limits; the \textsc{magphys} results do not change whether these are treated as non-detections or 1$\sigma$ fluxes.
 }
 \label{fig:magphys}
\end{figure}

\begin{deluxetable*}{ccccccccccc}
\tablecaption{Long-wavelength photometry used to calculate aggregate star formation properties. \label{table:agg}}
\tablehead{\colhead{Name} & \colhead{S$_{100\mu m}$} & \colhead{S$_{160\mu m}$} & \colhead{S$_{250\mu m}$} & \colhead{S$_{350\mu m}$} & \colhead{S$_{450\mu m}$} & \colhead{S$_{500\mu m}$} & \colhead{S$_{850\mu m}$} & \colhead{log L$_{\rm IR}$} & \colhead{SFR} & \colhead{$\tau_{\rm dep}$} \\
\colhead{} & \colhead{(mJy)} & \colhead{(mJy)} & \colhead{(mJy)} & \colhead{(mJy)} & \colhead{(mJy)} & \colhead{(mJy)} & \colhead{(mJy)} & \colhead{(M$_{\odot}$)} & \colhead{(M$_{\odot}$ yr$^{-1}$)} & \colhead{(Myr)}
}
\startdata
Core & 4.1$\pm$1.7 & 35.0$\pm$4.8 & 62.2$\pm$13.9 & 86.7$\pm$19.9 & 47.9$\pm$7.7 & 75.9$\pm$18.5 & 12.2$\pm$1.1 & 13.19$\pm$0.05 & 2300$^{+280}_{-250}$ & 340$^{+39}_{-43}$ \\
ID5 & 1.1$\pm$0.2 & 2.6$\pm$0.5 & 14.0$\pm$4.3 & 14.6$\pm$4.4 & 12.8$\pm$4.1 & 13.9$\pm$1.5 & 4.1$\pm$0.8 & 12.43$\pm$0.08 & 400$^{+80}_{-70}$ & 180$^{+30}_{-36}$ \\
450.58 & $<$1.5 & $<$4 & 11.3$\pm$2.2 & 15.57$\pm$2.7 & 15.3$\pm$4.1 & 14.5$\pm$3.1 & 5.3$\pm$0.8 & 12.40$\pm$0.11 & 370$^{+110}_{-80}$ & 885$^{+264}_{-210}$ \\
\enddata
\end{deluxetable*}

\subsection{SED Fitting}\label{sec:sed}
We fit optical/infrared spectral energy distributions (OIR SEDs) to those spectroscopic sources with matches within 1$\farcs$5, in the \citet{Laigle2016a} COSMOS photometric catalog. 
We do this using the updated form of the energy balance code \magphys\, \citep{daCunha2008a, daCunha2015a}\footnote{\url{http://www.iap.fr/magphys/download.html}}, which handles obscured galaxies with a wider range of star formation histories (SFH) appropriate for high-redshift galaxies with significant dust content. 
This code uses the \citet{Bruzual2003a} stellar population synthesis code to model spectral evolution, combined with the \citet{Charlot2000a} model to compute the total IR luminosity absorbed and re-radiated by dust in the interstellar medium. 
The high-redshift version of this code takes into account the fact that the copious amounts of dust in early star-forming galaxies are optically thick to the light in stellar birth clouds, such that the optical and IR parts of the SED are not produced in the same parts of a galaxy (i.e., the IR SED is not simply the difference between the attenuated and unattenuated optical spectra). 

We use the data only out to 24\,\um\, because the six central CO sources in Pointing 1 are detected as a single \herschel/SPIRE source and deblending it will introduce new uncertainties in a portion of the SED that should not directly affect the stellar mass estimate.
To verify this, we also test the SED fits including ALMA 1.1\,mm dust continuum and the less-confused 100\,\um\, and 160\,\um\, data, which resulted in much worse fits and derived stellar masses $<10\%$ different from the fits that use only the OIR data. 
This is likely due to the issue of spatial decoupling of dust and stellar light, and the attenuation is well constrained with the IR point at 24\,\um. 
Figure \ref{fig:magphys} shows the individual OIR SEDs for the Pointing 1 sources, which all have reduced $\chi^2 \lesssim 1$.

\subsubsection{AGN Content}\label{sec:agn}
We inspect the \magphys\, SEDs and COSMOS photometry for a few key features which may indicate significant AGN activity in any of the protocluster constituent galaxies. 
First, we use several criteria to check for a mid-infrared contribution from an AGN. 
We check for a rising slope in the mid-infrared SEDs (i.e. increasing luminosity with increasing IRAC wavelength), which would indicate the presence of warm ($>$100\,K) dust in an AGN torus, and apply the IRAC color selection criteria from \citet{Donley2012a}, finding that none of our sources meet the AGN classification. 
Similarly, we check for an 8\,\um\, excess, which is the ratio of the photometry at that observed wavelength to the fitted SED, and find that it is never more than unity, indicating that nearly all of the mid-infrared emission is likely arising from the stellar population in all galaxies and not warm dust continuum.

We also searched the X-ray point source catalogs \citep{Civano2016a} and found no counterparts to any of the CO-identified galaxies in the hard, soft, or combined bands. 
Because there is additionally no evidence for obscured AGN in the mid-infrared SEDs, we do not see evidence for any Compton-thick AGN like that found in the $z=4$ DRC \citep{Vito2020a}.

Finally, we check for radio-loud AGN by measuring FIR-radio correlation using the 1.4\,GHz \citep{Schinnerer2007a} and 3\,GHz \citep{Smolcic2017a} COSMOS catalogs.
IDs 1, 3, 5, and 7 are detected at $>5\sigma$ at 3\,GHz. 
Only two sources (IDs 1 and 3) are also identified at 1.4\,GHz at $>5\sigma$, and for these we measure a radio spectral slope $\alpha = -0.73 \pm 0.11$ and $-1.41 \pm 0.24$ respectively; otherwise we assume $\alpha=-0.8$ for non-detections in the following calculations. 
We calculate the radio-FIR correlation, $q_{\rm IR}$, from Equation 2 from \citet{Delhaize2017a}, using the rest-frame 1.4 GHz luminosities and assuming that the total L$_{IR}$ (Table \ref{table:agg}) can be divided according to each source's fractional contribution to the total ALMA continuum flux.
The average $q_{\rm IR}$ is 2.16 $\pm$ 0.36.
For ID1 and ID3, the only galaxies which are detected at 1.4\,GHz, we find explicitly $q_{\rm IR}$ = 2.5 $\pm$ 0.4 and 1.5 $\pm$ 0.2 respectively. 
Following \citet{Delhaize2017a}, who find that $q_{\rm IR}$ decreases shallowly with redshift, we expect $q_{\rm IR} = 2.27\pm0.01$ for star-forming galaxies at $z=2.5$, which is consistent with our findings given our large errors.

If we assume that the canonical value of $q_{\rm IR}$ means that all radio flux is due to star formation, a lower $q_{\rm IR}$ than expected implies a radio flux excess from AGN (i.e., $L_{\rm 1.4 GHz} = L_{\rm SF} + L_{\rm AGN}$).
ID3 is consistent with a significant radio excess considering the error and its unusually steep $\alpha$, as \citet{Delhaize2017a} also predict $q_{\rm IR} = 1.65\pm0.01$ for radio AGN at $z=2.5$.
Further, ID3 has the largest CO line FWHM (1300 $\pm$ 140\kms) which is kinematically indicative of AGN or merger activity, even when compared to the typically broad ($\sim$600--800\,\kms) CO lines in high-redshift DSFGs \citep[e.g.,][]{Narayanan2009a}. 
Therefore, we classify ID3 as a low-luminosity radio AGN but this is unlikely to affect our measurements of the stellar masses or star formation rates.

\begin{figure}
\includegraphics[width=1.0\columnwidth]{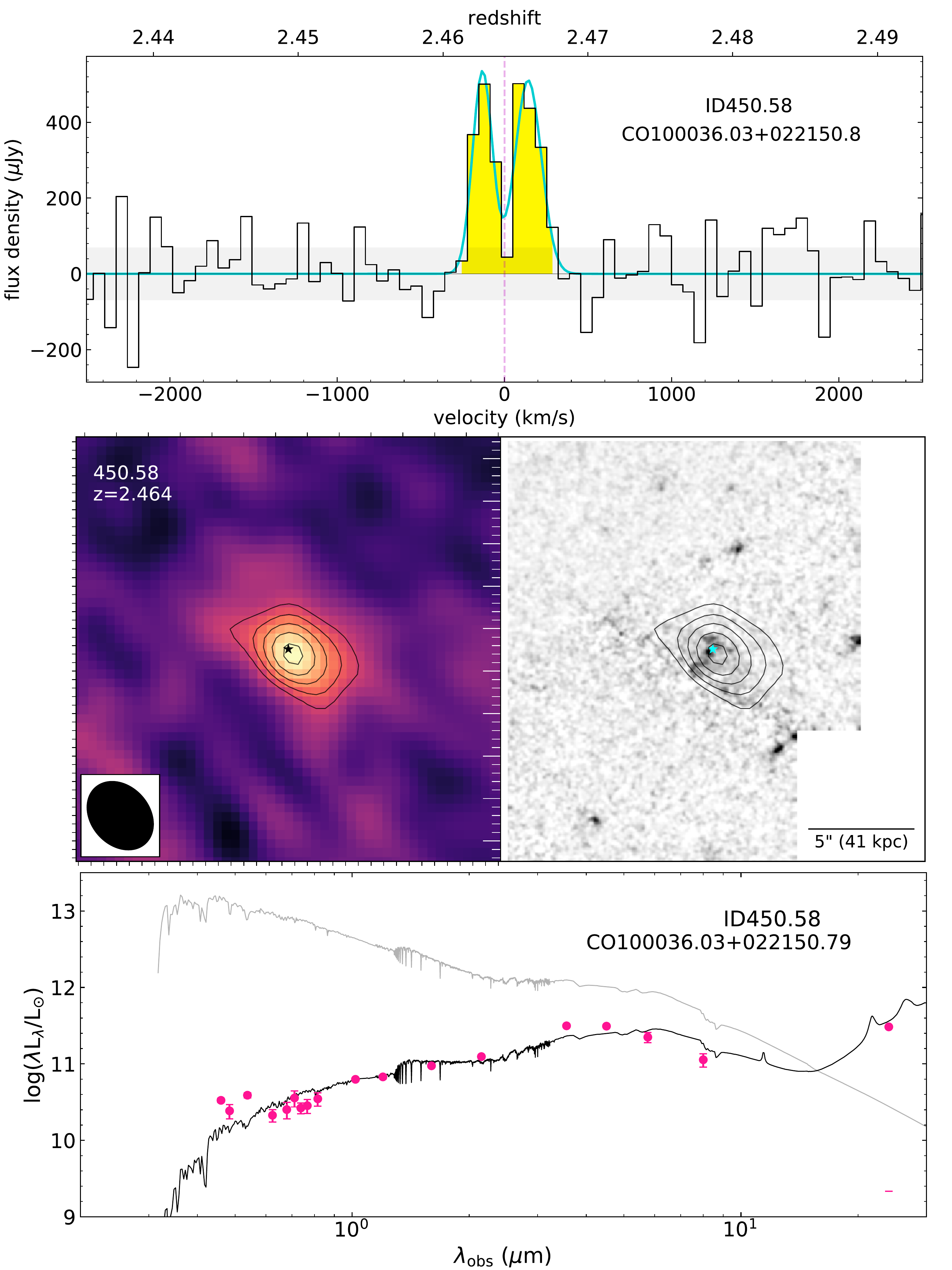}
\caption{Analysis for source 450.58 in Pointing 2.
\textit{Top:} CO(1-0) spectrum, presented the same way as Figure \ref{fig:spectra}. 
\textit{Middle left:} moment-0 map integrated over the same channels as highlighted in the spectrum; contours begin at 2$\sigma$ and increase in steps of $\sigma$, with the VLA beam shown in bottom left corner. 
\textit{Middle right:} $HST$ F814W image with VLA contours overlaid. 
Bright stars have been masked out. 
\textit{Bottom:} OIR SED as fit by \magphys, with symbols the same as in Figure \ref{fig:magphys}.}
\label{fig:45058}
\end{figure}

\subsubsection{Stellar Masses}\label{sec:stellarmass}
Since dust-obscured systems suffer heavy uncertainties due to degeneracies in reddening effects, geometry of star-forming regions, and reliability of OIR photometry, we verify the stellar masses using several methods.

As a first pass, we simply assume a constant mass-to-light ratio arising from the rest-frame near-infrared luminosity, using the methods outlined in \citet{Borys2003a} and \citet{Hainline2011a}. 
These studies model the M/L$_H$ and M/L$_K$ ratios using a sample of $z\sim2$ DSFGs fitted with the \textsc{Hyper-z} SED code \citep{Bolzonella2000a}, averaging between the most bursty vs. most continuous star formation histories.
$K$-band light has the advantage of having low sensitivity to previous star formation histories and dust obscuration, the main source of degeneracy for dust-obscured systems. 
$H$-band similarly traces redder stellar populations and suffers minimally from extinction from dust, and further minimizes the influence of thermally pulsating asymptotic giant branch stars \citep{Hainline2011a}.

$H-$ and $K-$band are both safe options here, with the main sources of uncertainty coming from the SED itself, possible contribution from an AGN due to the power-law emission of hot dust, and a lack of constraint on the dust attenuation in the IR regime. 
\citet{Hainline2011a} notes that uncertainties in the mass-to-light ratio due only to star formation history alone result in, at most, a factor of 2--3 difference in the stellar mass, but they cannot be constrained further without spatially resolved observations of the unobscured stellar populations (which will become possible with the \textit{James Webb Space Telescope}). 

A more constraining measurement of the stellar masses is the value computed from our own SED modeling using \magphys, whose inferred mass-to-light ratios have agreed with synthetic SEDs from simulations \citep[e.g.,][]{Hayward2015a}. 
We do find that for some galaxies, the measured stellar masses seem unphysically high given the galaxies' maximum possible ages, especially when compared with the observed stellar mass functions that are typically used for abundance matching \citep[e.g.,][]{Behroozi2010a}. 
However, this may not be surprising as other studies have found that dusty star-forming galaxies tend to dominate the high-mass end (log\,$(M_{\star}/M_{\odot}) > 10.3$) of the stellar mass function at $z>2$ \citep{Martis2016a}. 

We find that the stellar masses calculated from the rest-frame $H$-band luminosity are on the order of unity or slightly lower than the \magphys\, estimates, while the $K$-band estimates are either at unity or factors of 3--4 higher. 
It is possible there is a non-stellar contribution to the rest-frame $K$-band luminosity, but since we have checked for an excess in the mid-infrared photometry above the SED and did not find significant differences, this could be due to the two bands tracing stellar populations of slightly different ages. 
We use the \magphys\, estimates as the reported stellar masses, with the uncertainties given as a range between the lowest-mass and highest-mass results of all three methods. 
The stellar masses are fairly high, on the order of $\sim10^{11}$\,\msun\,, implying that these galaxies are mature and nearing their stellar mass limits given the age of the Universe at $z=2.5$. 
The $H$- and $K$-band photometry as well as the final stellar mass estimates are listed in Table \ref{table:phot}. 
In Section \ref{sec:discussion} we discuss the molecular gas fractions and the implication that these galaxies may be nearly quenched.

\subsection{Star Formation Rates}\label{sec:firsed}
To calculate the star formation rates and L$_{\rm FIR}$, we model the far-IR SED as outlined in \citet{Casey2012c}, using $Herschel$/PACS 100\,\um\, and 160\,\um\, as well as SPIRE 250, 350 and 500\,\um\, photometry. 
Because of the poor angular resolution of SPIRE (FWHM 36\arcsec\, at 500\,\um), six of the sources in Pointing 1 are spatially confused as one source, with an additional independent detection corresponding to ID5. 
We do not deblend the confused source due to deblending model degeneracies. Since we are interested in the aggregate quantities for the protocluster core we can still evaluate the total depletion timescale using the total star formation rate.
Thus, the star formation rate, far-infrared luminosity, and gas depletion timescales are calculated individually for ID5 and 450.58, and as aggregate quantities for the six core galaxies as well as any galaxies that may lie at the same redshift but are not detected in CO. 
After extracting the \herschel\, photometry, we fit a modified blackbody with a fixed slope ($\beta$ = 1.8) where only $\lambda_{\rm peak}$ is allowed to vary.
This results in a total SFR for the six core galaxies of 2300$^{+280}_{-250}$ \sfr, 400$^{+80}_{-70}$ \sfr\, in ID5, and 370$^{+110}_{-80}$ \sfr\, in 450.58.
The relevant photometry and IR-derived quantities are listed in Table \ref{table:agg}. 

\begin{figure*}
    \gridline{\includegraphics[width=1.0\columnwidth]{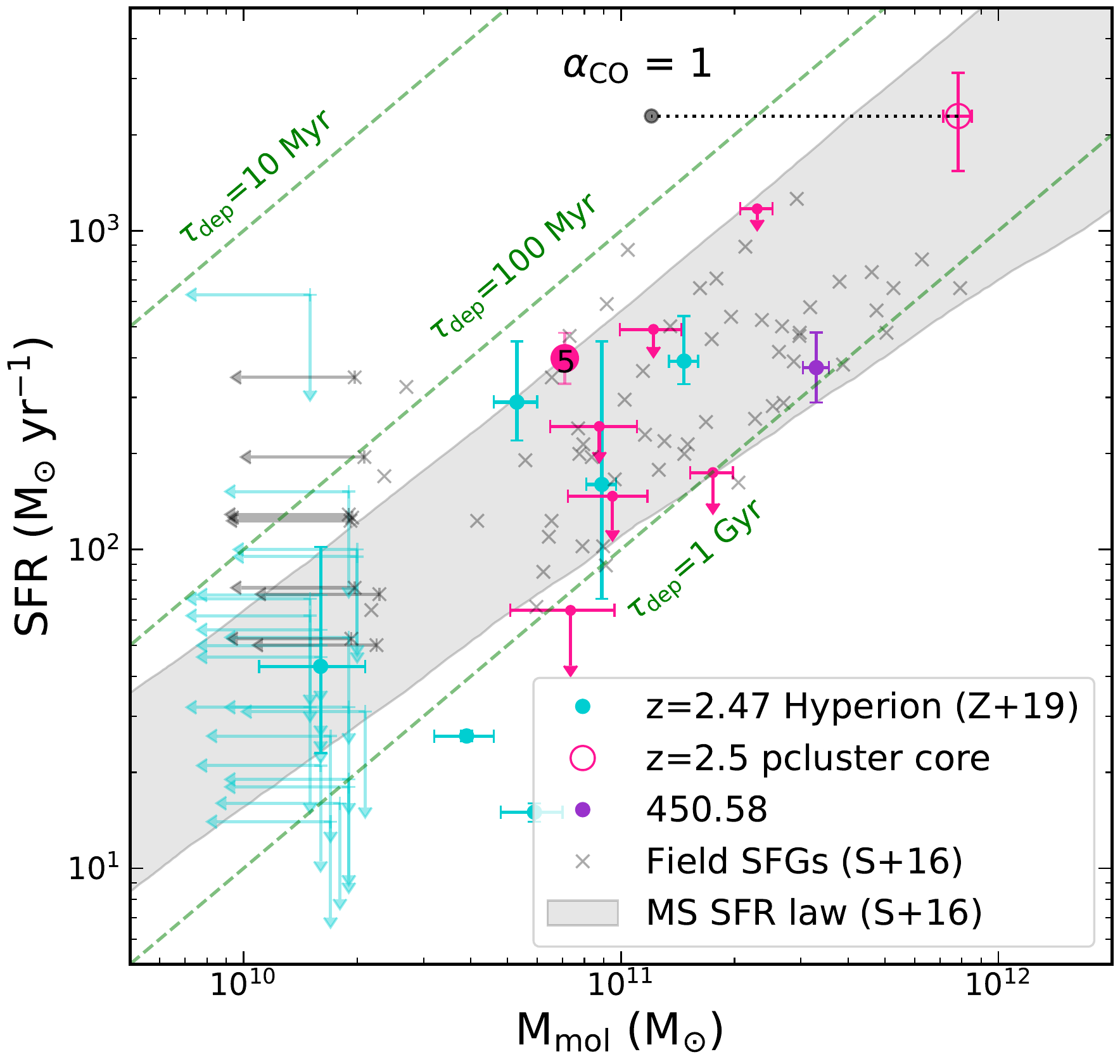}
    \includegraphics[width=1.0\columnwidth]{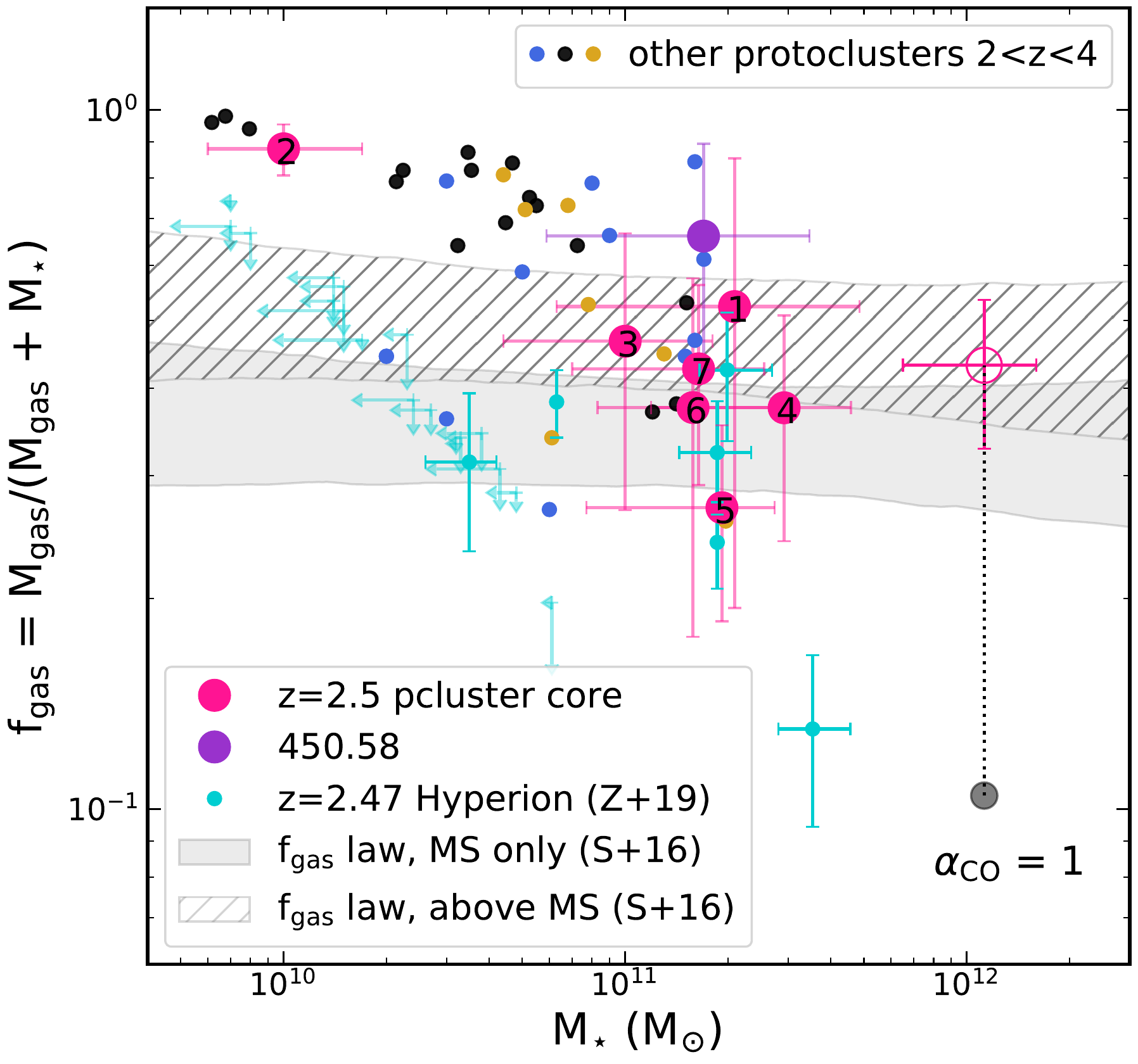}}
        \caption{\textit{Left:} star formation efficiency, measured by SFR vs. molecular gas mass for the $z=2.5$ protocluster core (empty pink circle), ID5 (filled pink circle), 450.58 (purple circle), the $z=2.47$ protocluster \citep[cyan;][]{Zavala2019a}, and field galaxies \citep[grey crosses;][]{Scoville2016a}. 
        The shaded region shows the parameterized SFR law for main sequence galaxies as in \citet{Scoville2016a} which is evaluated at $\langle z \rangle=2.2$. 
        The pink dots are the six individual galaxies which make up the inner core (empty pink circle), if we assume their fractional contributions to the total $L_{\rm IR}$ are distributed according to their ALMA fluxes. 
        Constant gas depletion times of 10\,Myr, 100\,Myr, and 1\,Gyr are shown in dashed green.
        \textit{Right:} Gas fraction as a function of stellar mass, with same symbols as left. 
        The shaded region shows the $\langle z \rangle=2.2$ scaling relation for main sequence galaxies from \citet{Scoville2016a}, while the hatched region shows that for galaxies above the main sequence.
        We also compare to the $z=4$ DRC \citep[blue;][]{Long2020a}, $z=2.16, z=2.49$ and $z=2.53$ protoclusters \citep[black;][]{Tadaki2019a}, and $z=2.53$ 4C23.56 \citep[gold;][]{Lee2017a}.
        In both panels, the dotted black line shows how our quantities change if one assumes \alphaco\, = 1 rather than 6.5 \acounits.}
    \label{fig:sfr_ms}
\end{figure*}

\section{Physical States of Member Galaxies}\label{sec:physstate}
Regardless of whether the galaxies in these two pointings are bound to a single halo within the parent protocluster Hyperion, we can measure the degree to which they may be the ``red and dead" members typical of relaxed clusters. 
One way to do this is by using the rest-frame $U-V$ and $V-J$ colors to distinguish between quiescent and star-forming galaxies. 
We find that all of these galaxies but one fall within the star-forming regime \citep{Muzzin2013a, vanderBurg2013a}, suggesting that these galaxies are not part of a red sequence. 
The only galaxy in the quiescent regime is ID3, which we claimed in Section \ref{sec:agn} to be a low-luminosity radio AGN. 
This further supports the scenario that ID3 is a quiescent galaxy in a post-starburst phase.

Two other common measurements to diagnose the current star formation mode are the gas depletion time scale and the fraction of gas mass relative to the stellar mass.
The gas depletion timescale ($\tau$ = M$_{\rm gas}$/SFR, or the inverse of the star formation efficiency) does not necessarily indicate a timescale for quenching, since there are likely to be changes in the gas mass due to non-star-forming processes (e.g. depletion via ram pressure stripping from the ICM or AGN outflows, or conversely accretion via instreams of gas from filaments). 
However, if protocluster galaxies undergo rapid bursts of extreme star formation, this would be reflected by very short depletion timescales through either prolific star formation rates or very small molecular gas reservoirs. 
We note that both of these measurements, however, are directly dependent on the assumptions one uses for the conversion factor \alphaco.
We find here that using \alphaco = 1\,\acounits\,, rather than 6.5\,\acounits\, which has been used throughout this paper (see Appendix \ref{app:aco}), results in an extreme departure from the typical ISM conditions in most galaxies --- notably, it results in gas-to-dust ratios of $<$20 \citep[4--5 times lower than the typical GDR in solar-metallicity galaxies regardless of redshift;][]{Remy-Ruyer2014a}. 

The total depletion timescale for the six core galaxies is 340 $\pm$ 40\,Myr, 180 $\pm$ 35\,Myr for ID5, and885 $\pm$ 250 Myr for 450.58.
If all of the molecular gas is consumed by star formation, we find that the galaxies in this protocluster can sustain their current star formation rates until $z=2.2$, at which time they will have built up stellar masses of log\,(M$_{\star}$/M$_{\odot}$) $\approx$ 11.0--11.7, which approach the maximum stellar masses seen in members of even the richest clusters at $z\sim1$ \citep{vanderBurg2013a}. 
Indeed, this supports the idea that intervening quenching mechanisms and shock heating of fresh gas likely occur before all of the molecular gas is converted into stars. 
Nonetheless, a long implied depletion timescale generally rules out bursty, stochastic star formation. 

Generally, protocluster studies at a variety of redshifts ($2<z<5$) have found that the constituent galaxies have relatively long gas depletion timescales with mixed results regarding the effect of environment. 
For example, in the DRC protocluster at $z=4.002$, \citet{Long2020a} found depletion times of $\sim$300\,Myr, expected for the field at that redshift. 
\citet{Miller2018a} also found a depletion timescale of $\sim$500\,Myr (correcting for \alphaco) in SPT2349-56 at $z=4.3$. 
At lower redshifts, \citet{Emonts2018a} found a longer depletion timescale of 1.5\,Gyr for the central radio galaxy in the Spiderweb protocluster at $z=2.16$, implying that overdense environments do not significantly impact constituent galaxy evolution.
Unlike the range of depletion timescales we find here, however, \citet{Tadaki2019a} found that the gas depletion timescales of individual galaxies in three protoclusters at $z\sim2$ was a function of stellar mass, where galaxies with log$M_{\star} > 11.0$ had a rapid $\tau_{\rm dep} < 500$\,Myr, but galaxies with lower stellar mass had longer depletion timescales than the field scaling relations ($>$1\,Gyr), implying mass-dependent effects of the dense environment.
We conclude that the core does not display evidence for enhanced star formation efficiency (or accelerated gas depletion), as was also found for a COSMOS protocluster at $z=2.10$ \citep{Zavala2019a}.
These galaxies are approaching their maximum stellar masses but they are not yet quenched, and are likely still undergoing interactions.

Figure \ref{fig:sfr_ms} directly compares the SFRs and gas masses of the CO sources as a proxy for star formation efficiency. 
In addition to the total SFR calculated in Section \ref{sec:firsed}, we also plot individual SFRs by assuming the fractional contribution of each galaxy to the total $L_{\rm IR}$ scales as their contribution to the total 1.2\,mm flux (however, these are upper limits on the individual SFRs, since the total \herschel\, flux likely includes line-of-sight interlopers and faint galaxies at the same redshift). 
Compared to field studies of main sequence submillimeter galaxies at $z\approx2$ \citep[and applying the same \alphaco, e.g.][]{Scoville2016a}, these galaxies do not individually show any evidence for enhanced star formation efficiency. 
\citet{Wang2018a} also studied the Pointing 1 core presented in this work, and using $L_{IR}/L'_{CO}$ they find a diversity of star formation efficiencies, with protocluster member galaxies both above and below the field scaling relations.
The sources that overlap with this work appear to have more enhanced SFE than we find, although the two studies use different methods: \citet{Wang2018a} uses the individual 24\um\, fluxes to extrapolate $L_{IR}$ and assume \alphaco=4 to measure gas masses, both of which result in a higher SFE.
The aggregate star formation rate of 2300\,$M_{\odot}$yr$^{-1}$ is quite high, but is not unreasonable for a collection of dust- and gas-rich galaxies in overdense environments.

Similarly, the gas fractions are in line with the field scaling relations: while these core galaxies have had time to build up impressive stellar masses in excess of $10^{11} M_{\odot}$ each, they are still gas-rich enough to continue forming stars and do not show evidence of quenching one may expect from a relaxed, mature galaxy cluster. 
In Figure \ref{fig:sfr_ms} we can see that the galaxies are consistent with main sequence galaxies or slight enhancement of the gas fraction above the main sequence.
This finding is consistent with the conclusions of another protocluster study at $z\approx2$ \citep{Tadaki2019a}, which was that gas fractions are enhanced at stellar masses $\rm logM_{\star} < 11.0$ but resemble main sequence scaling relations at higher stellar masses.
\citet{Lee2017a} finds the same trend for 4C23.56 at $z=2.49$: the gas fraction decreases as a function of stellar mass, falling in line with main sequence scaling relations at the same log$M_{\star}$ = 11.0 boundary.
This protocluster appears consistent with this finding: given the large errors, on the stellar mass, these galaxies imply slightly enhanced or normal gas fractions at log$M_{\star} \approx 11$, suggesting that gas accretion could be mass dependent.
Figure \ref{fig:sfr_ms} also shows how this conclusion changes if one assumes a lower value of \alphaco, which implies moderately enhanced star-formation efficiencies and unusually low gas fractions, compared with both the field and other protocluster galaxies at $z=2.10$ and $z=2.47$ \citep{Zavala2019a}.

\section{The Parent Structure at $z=2.47$} \label{sec:discussion}

\begin{figure*}
\begin{center}
    \includegraphics[width=2.0\columnwidth]{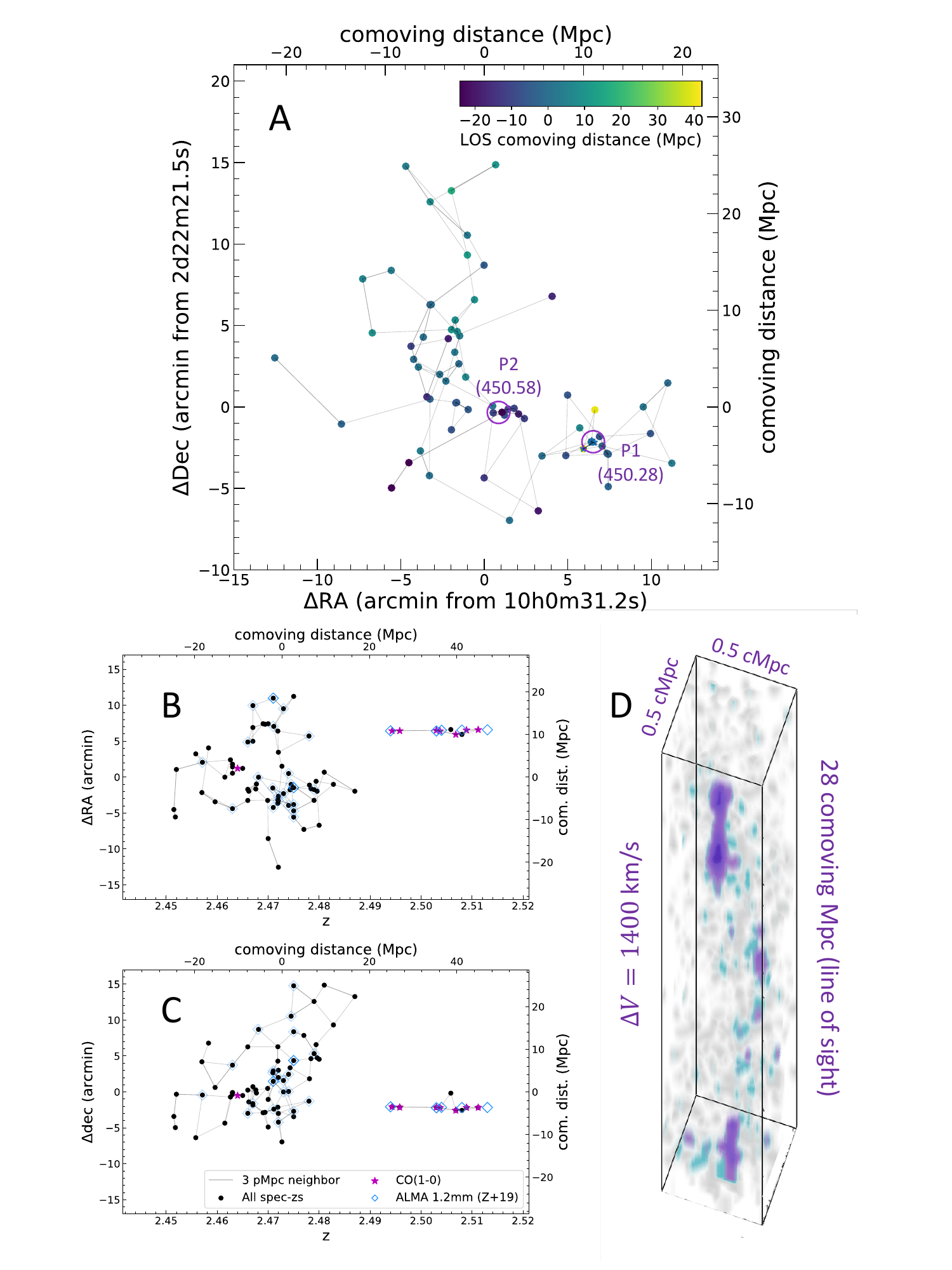}
    \caption{\textit{Panels A, B, and C:} all confirmed members of the $2.45<z<2.52$ extended structure (circles), linked to its two nearest neighbors with FoF algorithm with a 3 proper Mpc distance threshold. 
    Blue diamonds show ALMA Band 6 detections from \citet{Zavala2019a}, with thinner squares for detections $<3\sigma$ (note that redshifts are based on \citet{Wang2016a} so they are slightly offset from the CO sources reported here). 
    Magenta stars show CO detections from this paper. 
    \textit{Panel D:} 3D distribution of CO gas in the center of Pointing 1 (excludes ID5). 
    The VLA data is clipped at 2$\sigma$ (grey) and increases in steps of 1$\sigma$ (blue, lilac, dark purple).}
    \label{fig:3d}
   \end{center}
\end{figure*}

To place the two VLA pointings in context with the parent structure at $z=2.47$, we performed a friends-of-friends analysis \citep{Huchra1982a} on the eight detected galaxies and the previously spectroscopically confirmed members of the Hyperion super-protocluster \citep{Cucciati2018a}, which contains a diverse population of constituent galaxies from low-mass LAEs to extreme DSFG systems.
Hyperion includes 78 spectroscopic members from the $z=2.44$ LAE overdensity from \citet{Chiang2015a}, the $z=2.45$ overdensity in zCOSMOS from \citet{Diener2015a}, and the extended DSFG overdensity at $z=2.47$ from \citet{Casey2015a} (all shown in Figure \ref{fig:zhist}), plus the eight detections from this work for a total of 86 galaxies. 
We used a neighbor distance threshold of 3\,proper Mpc, in line with the expected linking lengths of unvirialized protoclusters at high redshift \citep{Chiang2013a}. 

Figure \ref{fig:3d} shows a 3D representation of the 74 spectroscopic members that pass the maximum friends-of-friends distance threshold, meaning that they are likely gravitationally bound at $z=2.47$.
Of the 12 Hyperion members that do not pass our criteria, several make up a $z=2.44-2.45$ filament which is not detected as a ``friend" of the $z=2.47$ overdensity.
The remaining $z=2.47$ superstructure plus the $z=2.5$ core discussed here span a staggering 60 comoving Mpc. 
Galaxy 450.58 was the only galaxy detected in CO in VLA Pointing 2 (indicated by the thick circles in Figure \ref{fig:3d}), but there are several other spectroscopic members of the protocluster that reside within 5 comoving Mpc, forming a local overdensity that is clearly part of the massive protocluster. 
On the other hand, the richly star-forming core of Pointing 1 is at a slightly higher redshift than the $z=2.47$ structure, located approximately 20 comoving Mpc from the center of the protocluster and thus not formally selected as a friend of the larger structure, but will likely be gravitationally bound to the $z=2.47$ structure by $z=0$ given its mass and proximity. 
This has interesting implications for whether or not this higher-redshift ``core" is a separate, independent galaxy cluster, or is in the process of falling into the parent structure.
The final panel in Figure \ref{fig:3d} shows the CO(1-0) 3D data cube extended along the $z$ axis, showing that it may be either a filamentary structure or a separate halo depending on the interpretation of the line-of-sight nature of the data.
In the following sections we discuss potential formation pathways for these two regions in context with the parent structure.

\section{Pointing 1: A protocluster core?}\label{sec:pointing1}

\subsection{Two Formation Hypotheses}
We dig further into the suggestion in the literature by \citet{Wang2016a} that the Pointing 1 core is a virialized cluster itself, possibly embedded within the larger protocluster. 
To do this we present two constrasting hypotheses. 
Under ``Hypothesis A," the galaxies in VLA Pointing 1 make up a virialized, nascent cluster core. 
In this scenario, the galaxies are likely occupying a spherical volume with the different redshifts traced to line-of-sight peculiar velocities. 
Under ``Hypothesis B," the galaxies of Pointing 1 may be extended in a filament along the line of sight and not yet virialized in a core. 
In this section we explore the evidence for both possibilities using the reported extended X-ray detection, our analysis of the member galaxies' dynamical states, estimates of allowable dark matter halo masses, and the relative rarity of a halo of this mass. 

A three-dimensional representation of the $z=2.503$ knot is shown in the fourth panel of Figure \ref{fig:3d}. 
Six sources in Pointing 1 span 15\,$\arcsec$ (140 physical kpc, or 490 comoving kpc) on the sky (with ID5 45\arcsec\, away) and span $2.494<z<2.511$ in redshift. 
Under Hypothesis A, this $\Delta z$ corresponds to a peculiar velocity range of $\Delta v = 1400$\,\kms, assuming the galaxies are spherically distributed in one halo. 
In this case, the gas would not be extended along the line of sight as suggested by Figure \ref{fig:3d}, making the spatial distribution of the total molecular gas significantly less complicated. 
We compared the galaxies' offsets in velocity from the center with what would be expected from a point source halo or an NFW halo with concentration $c=5$ \citep{Prada2012a}. 
The velocity distribution is not Gaussian, but the velocities are consistent with a $\sim$\,10$^{13}$\,\msun\, halo. 
It is possible we are sparsely sampling a population of galaxies that are not detected in this study, since the core contains only six galaxies, so a single bound halo is not immediately ruled out.

Under Hypothesis B, we assume the redshifts instead correspond to a line-of-sight filament, which has a maximum end-to-end size of 28 comoving Mpc (8 proper Mpc).
Figure \ref{fig:3d}, plotted as an extended filament, and shows that the 3D distribution of the sources is patchy and non-spherical. 
As discussed in the source extraction procedure (Section \ref{sec:VLA}), the peak flux centroids were straightforward to identify, but the edges of the molecular gas reservoirs are more ambiguous. 
The VLA data has a resolution of 2.5\arcsec\, ($\sim$20\,kpc) so we cannot meaningfully resolve mergers or outflows, but the diversity of line profiles and close separations imply these galaxies are not simple rotational disks. 
While we do not directly claim any of the CO detections to be formal mergers (as their optical counterparts are well resolved), it is clear from the extended nature of the overall molecular gas distribution that there is some gas interaction between galaxies. 
Under this hypothesis, we do not consider the galaxies to be relaxed.

\subsection{Extended X-ray signal: ICM or AGN?}\label{sec:icghost}
\begin{figure}
    \includegraphics[width=1.0\columnwidth]{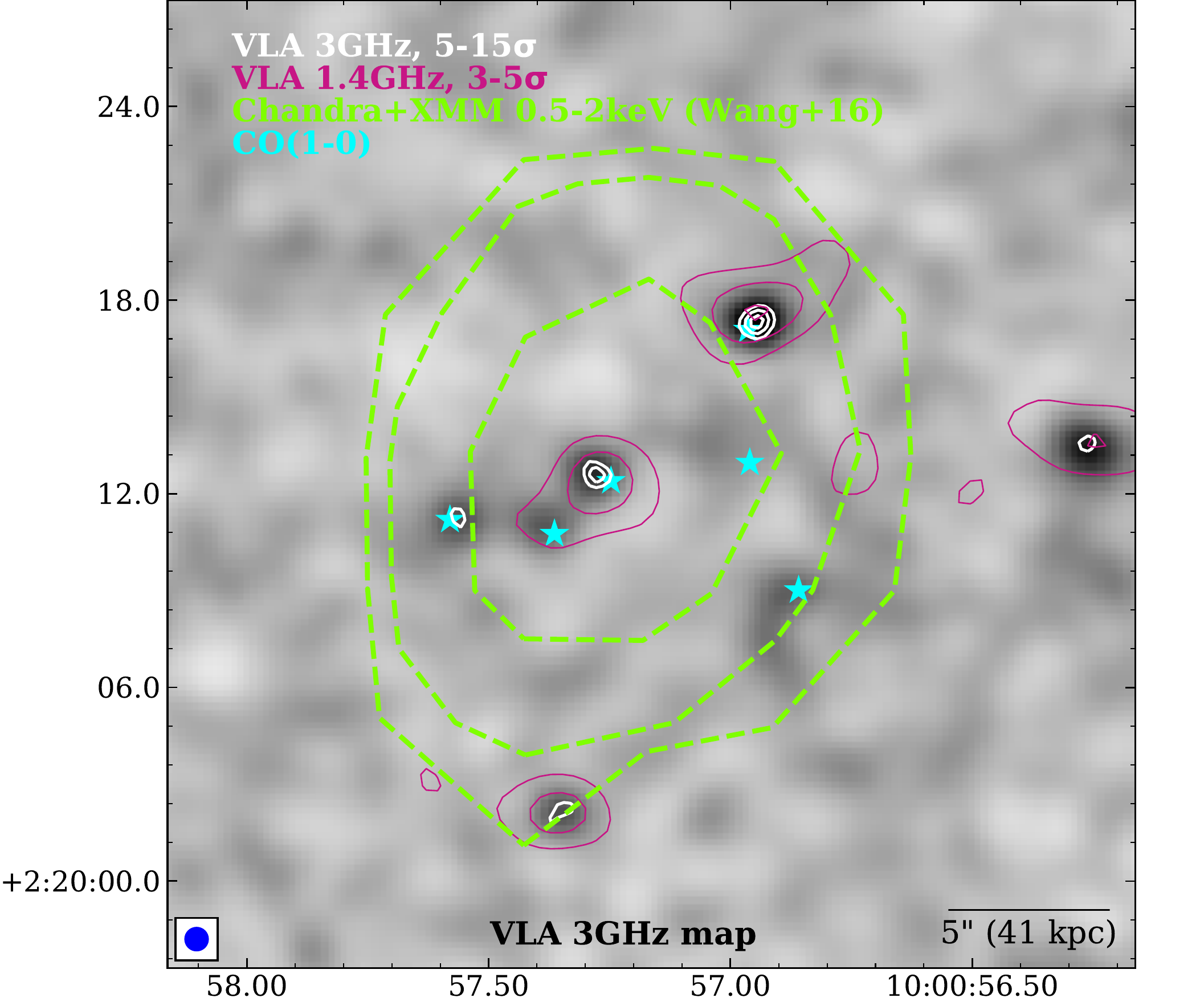}
    \caption{Background: 25\arcsec$\times$25\arcsec\, cutout of the VLA 3\,GHz COSMOS map. 
    The cyan stars indicate the centroid positions of the central six CO(1-0) detections in Pointing 1.  
    White: 3\,GHz continuum contours at 5-15$\sigma$ in steps of 3$\sigma$ \citep{Smolcic2017a}. 
    Pink: 1.4\,GHz contours at 3, 4, and 5$\sigma$ \citep{Schinnerer2007a}.
    Green: Smoothed \textit{Chandra+XMM} 0.5-2\,keV contours from \citet{Wang2016a}.}
    \label{fig:xray}
\end{figure}

\citet{Wang2016a} report a 0.5--2.0\,keV detection coincident with the center of Pointing 1 in the stacked COSMOS \textit{Chandra/XMM} maps which is 4$\sigma$ on 16\arcsec\, scales and $\sim$3$\sigma$ at 32\arcsec. 
They report a flux within a 32\arcsec\, aperture of (6.9 $\pm$ 2.0) $\times$ 10$^{-16}$\,erg s$^{-1}$ cm$^{-2}$, which we use as reported in our following calculations. 
We evolve a 2\,keV X-ray spectrum typical of galaxy groups to estimate the K-correction \citep{Boehringer2010a} and calculate a total luminosity of $L_{\rm 0.1-2.4 keV} = (8.9 \pm 2.0) \times 10^{43}$\,erg s$^{-1}$. 
Because it is more extended than a point source arising from a single galaxy and its luminosity is of order $\sim$10$^{44}$\,erg\,s$^{-1}$, \citet{Wang2016a} conclude that this source is evidence for a hot ICM, indicating that the core is already a virialized cluster. 
This would make it the highest-redshift galaxy cluster known to-date, consistent with our Hypothesis A.

In the event that the core is not yet virialized (Hypothesis B), the X-ray source may have plausible alternate origins, e.g. it may arise from a recently extinguished radio galaxy somewhere within the line-of-sight filament. 
In this scenario, the extended X-ray emission would outlive the radio detection of the galaxy. We will explore a possible alternative explanation of a so-called Inverse Compton (IC) ghost in greater detail.

Based on the higher-resolution \textit{Chandra} detection we can rule out a point source, so if real it is likely extended (albeit with low SNR), though there is the possibility that the X-ray detection is not a real source at all.
For the sake of arguing each hypothesis, we assume for now that the X-ray emission is real, extended, and associated with the protocluster core, as chance alignments of X-ray and radio emission in COSMOS have a probability of $<10^{-4}$ \citep{Jelic2012a}. 
Figure \ref{fig:xray} shows the \textit{Chandra/XMM} 2--3$\sigma$ contours reported by \citet{Wang2016a} as well as the 1.4\,GHz and 3\,GHz radio continuum emission in the central six galaxies in Pointing 1. 
There are two point sources detected at 1.4\,GHz which are coincident with CO detections (ID1 and ID3) totaling S$_{1.4}\approx140\,\mu$Jy, which we will return to in our discussion of Hypothesis B. 

To further explore the X-ray emission in context of Hypothesis A, we use the correlation between $L_X$ and cluster halo masses \citep{Leauthaud2010a} to estimate the minimum luminosity of the ICM at $z=2.5$. 
As described later in Section \ref{sec:halomass}, we calculate a maximum allowable dynamical mass range of the core of log\,$M_h/M_{\odot}$=12.9--14.3. 
In this mass range, we can calculate the expected luminosity of an ICM-dominated X-ray signal, using the $L_X-M_h$ relationship from \citet{Leauthaud2010a} whch was calibrated on groups and clusters in COSMOS. 
We find an expected range of log($L_X$) = 44.4--45.7 erg\,s$^{-1}$, which is a factor of 3--50 times brighter than the observed X-ray luminosity here. 
However, it is possible that the formation of the ICM at $z=2.5$ is still in its nascent stages and would not be as bright as the intra-cluster gas in more mature, lower-redshift clusters, even if they have similar masses. 

To explore Hypothesis B, we turn to \citet{Fabian2009a}, which discusses an extended X-ray detection coincident with the giant elliptical galaxy HDF130 at $z=1.99$ \citep{Casey2009a}.
This X-ray source, with luminosity $L_{2-10 \rm keV} \approx 5.4 \times 10^{43}$ erg/s (comparable to our source), shows evidence for the Inverse Compton ghost effect.
Radio-loud AGN, when active, accelerate electrons to Lorentz factors $\gamma > 10^4$, which in turn undergo Inverse Compton scattering off the CMB. 
Lorentz factors of $\gamma \sim 10^4$ are required to generate the GHz-frequency synchrotron emission in the radio band, but electrons with $\gamma \sim 1000$ can up-scatter CMB photons to the X-ray bands.
The Compton cooling time is ten times longer at X-ray wavelengths than at radio wavelengths, so the ``ghost" of the extended lobes of the AGN would be detectable in X-ray long after the extended radio signal has died out. 

Electron acceleration ends when the AGN shuts off, so the higher-frequency radio emission arising from synchrotron radiation will die off first. 
This means that with a recently extinguished radio galaxy we would not necessarily detect a luminous ($\geq$ mJy flux densities) radio AGN counterpart to the X-ray emission at 1.4\,GHz \citep[see e.g.,][]{Simonescu2016a} but rather at MHz frequencies\footnote{While lower-frequency (325 MHz) radio observations exist for the COSMOS field \citep{Smolcic2014a}, this survey is too shallow to detect an extended (i.e., lobed) radio counterpart to ID3.}. 
Indeed, the 1.4\,GHz radio emission in HDF130 is present as point sources only at the 400\,$\mu$Jy level, more consistent with a low-luminosity AGN. 
The most obvious candidate analog of HDF130 within our Pointing 1 is ID3, which appears to be somewhat radio luminous given its submillimeter emission; thus, its 140\,$\mu$Jy flux at 1.4\,GHz could also be attributable to a low-luminosity AGN.

In terms of morphology, the double-lobed X-ray source in \citet{Fabian2009a} is much more linearly extended ($\sim$40\arcsec) than the single source considered here. 
On the most compact scales (10\arcsec), the morphology of the \citet{Wang2016a} source appears mostly spherical or slightly elongated in the NW direction, and does not show evidence for lobed structure. 
However, the detection beyond 20$\arcsec$ is marginal (2--3.3$\sigma$). 
In both models \citep[e.g.,][]{Mocz2011a} and observations \citep[e.g.,][]{Jelic2012a}, the X-ray IC lobes arising from powerful radio galaxies are expected to span a few hundred up to a thousand kpc, or $\sim$25--120\arcsec\, on the sky. 
The source here spans only 20\arcsec\, on the sky at maximum (170 proper kpc), on the smaller end of the range of IC ghost lobe sizes \citep[c.f. 690 kpc for HDF130 in][]{Fabian2009a}. 

The measured 20\arcsec\,size of this source also happens to be a factor of 2--3 smaller than the typical ICM diameter of 600--700\,kpc in low-redshift clusters \citep{Piffaretti2011a} of similar mass (see halo mass calculations in Section \ref{sec:halomass}), though it is unclear if the ICM would be physically smaller in a structure that has very recently collapsed. 

Finally, the relatively low signal to noise ratio of the X-ray signal at large radii causes some additional concern. 
For example, the \textit{Chandra/XMM} flux drops off quickly --- $<$50\% of its peak value at double the extraction radius --- which is not typical of the radial profiles of intra-cluster gas. 
However, the overall low signal to noise ratio of the extended \textit{Chandra/XMM} detection ($<$3$\sigma$ beyond 30\arcsec and $4\sigma$ at 16\arcsec) makes the radial profile difficult to constrain and the extended nature marginal. 
Though the \textit{XMM}-only map has better sensitivity, the source is several arcminutes off-axis in the COSMOS maps so it is a marginal point source detection as well. 
Based on its size and luminosity, we argue that the X-ray signal, if real, is more consistent with emission from a single galaxy IC ghost than with an ICM origin. 
In Section \ref{sec:prob}, we discuss the probabilities of observing either scenario (non-thermal IC emission or ICM gas) given the source's selection in the COSMOS survey field.

\begin{figure}
    \includegraphics[width=1.0\columnwidth]{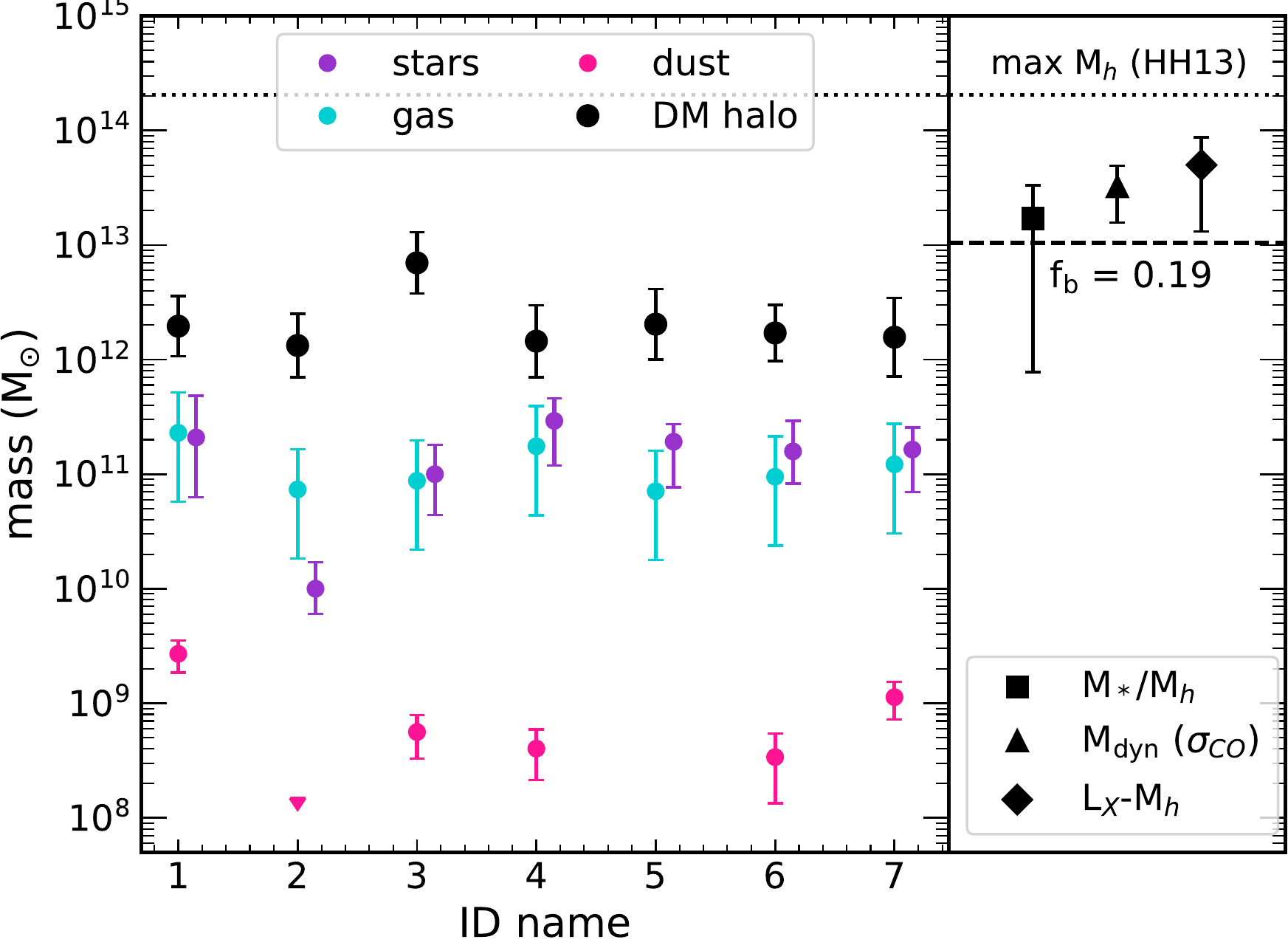}
    \caption{\textit{Left:} summary of all mass quantities calculated for the protocluster core in Pointing 1: black circles are halo masses calculated via abundance matching, purple circles show average stellar masses, cyan circles show molecular gas masses, and magenta points show dust masses calculated from ALMA Band 6 flux (note that ID2 is an upper limit, and there is no dust mass measurement for ID5). 
    \textit{Right:} summary of total dark matter halo calculations for Pointing 1. 
    Triangle shows total halo mass calculated from a dynamical mass using the CO(1-0) velocity dispersion; square shows the halo mass calculated from abundance matching (a sum of all individual halos); diamond shows a halo mass derived from the X-ray luminosity-mass ratio from \citet{Leauthaud2010a}. 
    The dotted line shows the 95th percentile maximum allowable halo mass at $z=2.5$ \citep{Harrison2013a} and the dashed line shows the minimum reasonable halo mass assuming a baryonic fraction of f$_b$ = 0.19.}
    \label{fig:allmasses}
\end{figure}

\subsection{Halo masses for the $z=2.5$ core structure}\label{sec:halomass}
Estimating the halo mass for this source necessarily depends on whether these galaxies occupy a single dark matter halo or each reside in individual halos at $z=2.5$, so we use several methods to calculate halo masses using different assumptions about the available CO, X-ray, and optical data. 
For each of the halo mass estimators we discuss, we also calculate the equivalent $z=0$ halo masses based on the halo assembly histories used by \citet{Harrison2013a} in their publicly available code to calculate relative halo rareness. 
These authors describe how to compute curves in mass-redshift space along which the number of collapsed halos at that mass and redshift are equally rare according to the selected rareness statistic. 
Thus, the equivalent $z=0$ halo mass is that which has the same value of the rareness statistic as the $z=2.5$ halo. 
We caution that the equivalent $z=0$ halo mass \emph{not} a prediction for the descendant mass of the halo --- there is significant scatter in the final halo masses of nascent protocluster halos \citep[e.g.,][]{Chiang2013a, Lovell2018a}, so we cannot constrain this further.

The following calculations assume Hypothesis A holds, in which the extended morphology of the molecular gas along the line of sight would be attributable to velocity dispersion of sources within a single gravitationally bound halo. 
We can place a hard lower limit on the single halo mass by assuming that the baryonic concentration within the halo is not below the cosmic baryonic fraction, $f_b$=0.19 \citep{Planck2015a, Behroozi2018a}, reaching a conservative minimum of log$(M_h/M_{\odot}) \approx 12.9$. 
A more useful lower limit for Hypothesis A, though, is the minimum halo mass for which the CO galaxies could be bound to a single halo. 
As in Section \ref{sec:physstate}, we compared the velocity offsets of each galaxy with those expected from an NFW halo. 
The minimum halo mass at which these galaxies can be considered bound is log$(M_h/M_{\odot}) \approx 13.2$. 
Although it is a small sample size, the core galaxies' velocities are consistent with a bound dark matter halo above this mass. 
This minimum halo mass at $z=2.5$ has an equivalent $z=0$ halo mass of approximately 8$\times$10$^{13}$\,\msun. 

Also under Hypothesis A, we estimate the dynamical mass (M$_{\rm dyn}$ = M$_{\rm halo}$ + M$_{\rm baryon}$) using the virial theorem:

\begin{equation}
    M_{\rm dyn} = C \frac{R \sigma^2}{G}
\end{equation}

\noindent where $R$ is the proper size of the structure, $\sigma$ is the velocity dispersion with respect to the center, and C is a constant taken to be 3 for spherical mass distributions \citep{Ho2000a}.  
We take $R$ to be 190\,kpc, which is the transverse size of the Pointing 1 core \citep[and consistent with the X-ray size of 130\,kpc reported in][]{Wang2016a}.
The velocity dispersion of the CO detections, $\sigma$, with respect to the average redshift of the seven sources ($z=2.503$) is measured to be $\sigma=500\pm120$\,\kms.  
We estimate M$_{h(z=2.5)}$ = (3.3 $\pm$ 0.6) $\times$ 10$^{13}$\,M$_{\odot}$, which translates to a $z=0$ halo mass of approximately 5$\times$10$^{14}$\,\msun, comparable to the Coma cluster. 
This is consistent with the dynamical mass estimate of $3.2^{+1.9}_{-1.2} \times 10^{13}$\,\msun\, from \citet{Wang2016a}, which was calculated using the transverse extent of the X-ray detection.

Still assuming Hypothesis A, we also calculate the halo mass directly using the luminosity of the 0.5--2\,keV X-ray detection assuming it arises from a hot ICM in a cluster-sized halo. 
We use the K-correction described in Section \ref{sec:icghost} to derive the rest-frame luminosity at 0.1--2.4\,keV and then use the $L_X/M_h$ relation from \citet{Leauthaud2010a}. 
This results in a halo mass estimate of M$_{h(z=2.5)}$ = 
(5.0 $\pm$ 0.2$) \times10^{13}$\,M$_{\odot}$ (or M$_{h(z=0)} \sim 9\times10^{14}$\,\msun) which is the highest mass estimate of any of our methods.
One should note, however, that this correlation with halo mass has been measured only out to $z=1$ for virialized cluster cores. In the event that there is a redshift evolution in this correlation, this halo mass would be inaccurate. 
 
Alternatively, under Hypothesis B, the redshifts of the CO sources may instead correspond to true separations along the line of sight and we assume they occupy completely separate dark matter halos.
Under this hypothesis, we calculate the total halo mass by summing the individual halo masses of the seven galaxies. 
To measure the individual halo masses we use abundance matching, using the stellar--halo mass relationships of \citet{Behroozi2010a}. 
Since our measurements of the stellar masses approach the upper limit of the predicted $z\sim2$ stellar masses according to \citet{Behroozi2010a}, this sometimes results in invalid extrapolations to compute halo masses.
To avoid this problem, we use a Monte Carlo approach assuming a log-normal distribution of stellar masses for each galaxy, centered on the reported stellar masses in Table \ref{table:phot}. 
Taken in aggregate, this results in a total halo mass of M$_{h (z=2.5)}$ = (1.7 $\pm$ 0.8) $\times 10^{13}$\,\msun. 
This should be taken as an upper limit for Hypothesis B because it represents the extreme scenario in which there is no gravitational interaction between any of the galaxies; here we cannot account for the possible double-counting of overlapping dark matter halos. 
If we assume that these galaxies --- currently extended in a filament --- will eventually collapse into a single halo, its equivalent halo mass at $z=0$ is $2\times10^{14}$\,\msun. 
While this is the smallest mass estimate of our three methods by a factor of a few, it still implies a Coma-sized cluster at $z=0$.

Finally, the absolute maximum allowed halo mass is found by calculating the exclusion curve as a function of redshift, i.e. the halo mass above which it would come into tension with our $\Lambda$CDM cosmological parameters. 
Using the halo mass assembly history from \citet{Harrison2013a}, we find that there is a $<5$\% chance of detecting a halo more massive than $2 \times 10^{14}$ \msun within the COSMOS survey at $z=2.5$, so we can safely assume this protocluster core is not more massive than this.

In Figure \ref{fig:allmasses} we show a summary of each of our baryonic and halo mass estimates. 
The allowable halo mass ranges from log\,$M_h/M_{\odot}$ = 12.9--14.3, where the minimum represents the lowest baryonic fraction and the maximum is that allowed by our cosmological assumptions. 
Within the errors, all of our halo mass estimates are consistent with one another: under either hypothesis, this is a uniquely massive overdensity which will rival the largest galaxy clusters by $z=0$. 
It is likely that the halo structure lies somewhere between the two extremes of a single coalesced halo and seven completely separate halos. 
Because the CO-detected galaxies are close enough (in the transverse direction, regardless of peculiar velocities) that there is probably overlap in their dark matter halos, it is likely, no matter the configuration at $z=2.5$, that the seven galaxies in Pointing 1 will collapse into a single halo at lower redshift. 
Indeed, it is plausible that they will form a single brightest cluster galaxy within the $z=2.47$ parent cluster by $z=0$. 

\begin{figure*}
\begin{center}
    \includegraphics[width=1.5\columnwidth]{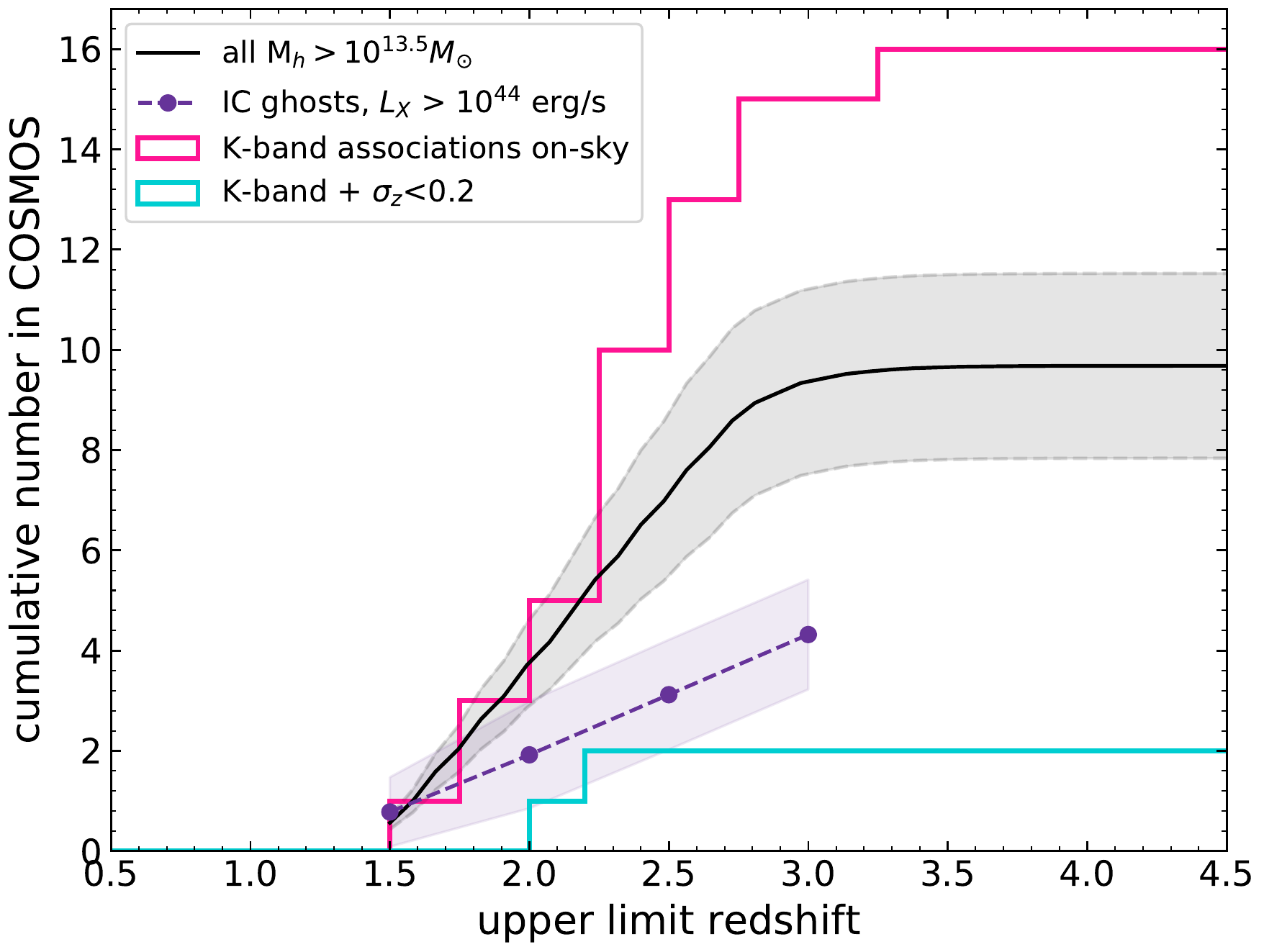}
 \end{center}
    \caption{Black curve: detectable number of extended X-ray halos with equivalent $z=0$ masses to the $z=2.5$ protocluster core in the 2 sq. degree COSMOS field, calculated using the framework of \citet{Harrison2013a} and assuming the halo mass is calculated from the X-ray luminosity. 
    Magenta histogram: the number of 2D (on-sky) associations of 10 $K$-band-selected (i.e., massive) galaxies actually observed in the COSMOS field. 
    Cyan histogram: the number of 3D (on-sky plus redshift) associations of 10 $K$-band-selected galaxies in COSMOS where $\sigma_z$ of the 10 galaxies is $< 0.2$. 
    Purple points: the expected number density of Inverse Compton ghosts with $L_X \geq 10^{44}$ erg/s  visible in the COSMOS field. 
    Approximately half of all halos as massive as the protocluster core at $z=2.5$ are expected to host massive radio galaxies which could give rise to IC ghosts, while we expect only 1--2 halos to host overdensities of at least 10 massive galaxies.}
    
    \label{fig:halorarity}
\end{figure*}

\subsection{Quantitative rarity of massive halos}\label{sec:prob}
In order to place the relative rarity of the $z=2.5$ core structure in context, we compare the two formation hypotheses from a probabilistic standpoint. 
To do this, we calculate the probabilities that structures like the core discussed in this paper would be reasonably observable in a survey size of 2 square degrees, here corresponding to the COSMOS field in which the source was found. 
First, we estimate how many total halos we expect to exist in the COSMOS field with similar or greater mass than the $z=2.5$ core in this study which could be detectable in X-ray emission. 
We compare the probabilities of the X-ray emission in each halo originating from hot ICM gas, non-thermal emission like Inverse Compton processes, or an AGN with radio and X-ray lobes. 
Independently, we then estimate the number of halos in COSMOS massive enough to host massive radio galaxies, a population which likely overlaps with the halos that could emit X-rays. 
Finally, we estimate simply the rareness of a spatial overdensity of massive star-forming galaxies (which may not necessarily reside in a single halo nor be associated with AGN or ICM features).

\subsubsection{Rarity of X-ray-emitting halos}\label{sec:icmrare}
We estimate the total number of massive halos as rare as the core discussed here expected to be observable within the COSMOS field as a function of redshift. 
Following the methods used in \citet{Marrone2018a} for SPT0311-58, we use the halo rarity framework of \citet{Harrison2013a} as discussed in Section \ref{sec:halomass} to estimate the relative rareness of this protocluster core.
By imposing a physically sensible redshift range and observational limits of a survey, one can then use this statistic to infer how many halos in that survey volume might collapse into cluster-sized halos\footnote{Note that we consider halos with similar characteristics only to the $z=2.5$ protocluster \emph{core}, not the larger Hyperion protocluster, which contains a number of overdensity peaks or individual filaments within $z=2.4-2.5$ \citep{Cucciati2018a}. 
While all galaxies within some membership radius may be considered part of Hyperion, Hyperion itself is unlikely to be a single halo at $z=2.47$ \citep[see predictions from simulations, e.g.,][]{Chiang2017a}, so we would not be comparing apples to apples with this rarity framework.} of similar or greater mass by $z=0$. 

The necessary observational constraints on the rareness statistic are the survey area, the redshift range in which to search for objects, and the selection function or the limiting halo mass as a function of redshift. 

We describe the observational selection function for such rare halos using X-ray emission; an implicit assumption of this calculation is that all similarly rare halos \textit{will} be detectable in X-ray emission above a given mass threshold as a function of redshift.
Whether or not this is a realistic expectation is discussed further below.
The survey area is the 2 square degrees covered by both \textit{XMM} and \textit{Chandra} in COSMOS. 
The volume over which we calculate the probability of observing a halo of any mass and redshift that is equally rare as this protocluster core is enclosed by $1.5<z<4$, motivated by the exclusion of mature clusters at $z<1.5$, and the assumption that a virialized, detectable ICM would not physically exist before $z=4$ \citep{Chiang2017a}, nor would enough radio galaxies have turned off to produce IC ghosts in X-rays \citep{Mocz2011a}. 
Finally, we define a limiting halo mass as a function of redshift. 
We assign this based on the X-ray flux limits in COSMOS such that we are selecting for X-ray detectability.
In other words, we assume that the X-ray luminosity scales with halo mass, regardless of whether the halo is actually producing X-rays.
Then, using the redshift-dependent K-correction function noted in Section \ref{sec:halomass}, we estimate the expected observed luminosity of an X-ray--emitting halo as a function of redshift. 
This is then matched to a halo mass using the $L_X$-$M_h$ relation of \citet{Leauthaud2010a} resulting in a limiting halo mass function set by the \textit{Chandra} flux limit in the COSMOS survey (which is $2.2\times10^{-16}$\,erg\,s$^{-1}$\,cm$^{-2}$). 

Based on the survey limits alone, we estimate $\sim$10 possible halos similar to this protocluster core that are hypothetically \emph{observable} via extended X-ray emission within in the COSMOS field between $1.5<z<4$. 
This implies that, based on its halo mass alone, this source is not unphysically rare, and rather it might be expected given the relatively large size of the COSMOS field.

One significant caveat of this calculation, however, is that not all massive halos at high redshift will actually generate a detectable X-ray signal, either since not all massive halos will host AGN with lobes, and/or protoclusters may not have accreted sufficient gas to generate thermal Bremsstrahlung emission from a hot ICM. 
We stress that this result of 10 similar halos is the maximum number of halos that might be massive enough to emit detectable X-rays --- it is based simply on the X-ray flux limit of COSMOS and is not a predicted luminosity function of cluster ICM. 
Indeed, observations suggest that cluster halos do not undergo shock heating to virial temperatures until $z\sim2$ \citep[see compilation of][Fig. 11]{Overzier2016a}; the implication of that work are that the number of detectable halos of similar rarity in COSMOS would be reduced to a total of 4 halos, and the $z=2.5$ structure would be categorically outside of the allowable redshift range.
Additionally, studies evolving the X-ray and radio luminosity functions to high redshift \citep{Celotti2004a, Finoguenov2010a, Mocz2011a} suggest that clusters at high redshift ($z>2$) and luminosities greater than a few times 10$^{43}$ erg s$^{-1}$ should be exceedingly rare, while, in contrast, extended X-ray emission arising from Fanaroff-Riley Type II \citep*[FRII;][]{Fanaroff1974a} galaxies will dominate the extended X-ray source luminosity function. 
The cumulative upper limit number of halos as a function of redshift is plotted in black in Figure \ref{fig:halorarity}.

\subsubsection{Rarity of massive radio galaxy halos}\label{sec:radiogal}
Separate from the X-ray-emitting halo rarity calculation, we next estimate the relative volume density of radio galaxies residing in massive halos at $z=2.5$ which could at some point give rise to extended X-ray emission from IC ghosts. 
We use abundance matching to scale the radio luminosity functions from \citet{Novak2017a} at $z=2.5$ to their host halo masses of similar abundance using the $z=2.5$ halo mass functions from \citet{Behroozi2019a}. 
The duty cycle of the luminous radio galaxies, which are the IC ghost progenitors, needs to be accounted for in this calculation; we adopt a duty cycle of 1\% representing the radio-loud phase of massive galaxies above radio luminosities $L_{\rm 1.4\,GHz}>10^{23}$\,W\,Hz$^{-1}$ \citep{Best2005a}. 
However, the IC ghost phase during which a massive halo may have a detectable extended X-ray component, is $\sim$10 times longer than the luminous radio galaxy phase (the latter of which may only last $\sim$10--100\,Myr); thus, in our scaling of the radio luminosity function to halo masses, we apply a total duty cycle of 10\%. 
We infer that at $z=2.5$, a $2\times10^{13}\,M_\odot$ halo is likely to host a $L_{\rm 1.4\,GHz}\sim10^{24.8}$\,W\,Hz$^{-1}$ radio galaxy capable of generating an IC ghost with a 10\% duty cycle\footnote{This luminosity is just shy of the FRI/FRII luminosity threshold of $\sim$10$^{25.4}$\,W\,Hz$^{-1}$ \citep{Ledlow1996a} and well within the range of luminous FRI systems that have $L_{\rm 1.4\,GHz}>10^{23}$\,W\,Hz$^{-1}$.}. 
At this mass and luminosity, we estimate a total of $\sim$10 such radio systems to exist in the COSMOS field between a redshift of $2<z<3$.
This agrees roughly with the total number of extended X-ray sources detectable in the COSMOS field overall (estimated independently in Section \ref{sec:icmrare}), rendering it a plausible origin for an arbitrary high-redshift extended X-ray source. 

\subsubsection{Probability of detecting IC ghosts}\label{sec:icghostrare}
Lastly, we directly evaluate the probability of detecting an IC ghost at any redshift, using an alternate framework than what is presented in Section \ref{sec:radiogal} for massive radio galaxies.
\citet*{Celotti2004a} examine the relative numbers of radio galaxies and clusters as a function of redshift, estimating that Inverse Compton ghosts may have a space density of $3\times10^{-8}\,\rm Mpc^{-3}$ at $z>1$. 
However, these early estimations of the abundance of IC ghosts typically converted the radio luminosity function to X-ray flux and evolved this in time -- but since this method is dependent on the radio flux limit, it does not account for the time period in which a source is visible in X-ray but not radio wavelengths.  

\citet{Mocz2011a} performed detailed modeling of the radio galaxies which give rise to X-ray lobes, evolving these galaxies from their active jet phase to their IC ghost (i.e., X-ray only) phase to their eventual ``dead lobe" phase in which neither X-ray nor radio is detectable. 
During their active phase ($\sim$100\,Myr), jets will produce both X-ray and radio emission, but once the jet turns off, the radio emission dies off quickly ($<$10\,Myr), while the ``ghost" phase can last up to ten times longer depending on the initial conditions of the jet. 
\citet{Mocz2011a} shows that the lifetime of the X-ray--detectable lobes decreases with higher redshifts due to evolving interaction with the CMB, and further that the abundance of any X-ray lobes is dependent on the birth function (i.e., when jet activity initiates) of FR\,II radio galaxies. 
\citet{Mocz2011a} therefore derives a framework to calculate the probability of observing an IC ghost at any redshift given the probability density functions of the ages of these sources. 

Using this framework, we have calculated the number of expected IC ghosts in COSMOS from $1.5<z<3$ in bins of $\Delta z=0.5$. 
Note that calculating this number beyond $z=3$ is highly uncertain since the radio luminosity function is poorly constrained, but regardless, the abundance of IC ghosts is expected to reach a maximum at $1.5<z<3$ during the peak quasar era \citep{Celotti2004a, Mocz2011a}. 
We find there should be about 4--5 IC ghosts at $1.5<z<3$ above the \textit{Chandra} X-ray flux limit in COSMOS. 
This overall estimate is whin a factor of 2 of what we estimate for IC ghosts in Section \ref{sec:radiogal}.
This is in agreement with the prediction from \citet{Jelic2012a} that there should be approximately one non-thermal source within COSMOS detectable in both X-ray and radio, since the X-ray--only phase is expected to outlive the radio+X-ray phase by at least a factor of a few \citep{Fabian2009a}. 
This implies that roughly half of the massive radio galaxy halos discussed in the previous section should give rise to observable IC ghosts at $1.5<z<3$.

\subsubsection{ICM or IC ghost: the more likely physical scenario?}
We find it is observationally and cosmologically possible to see approximately 10 halos in COSMOS at least as massive as this structure (at $z=0$) associated with extended X-ray detections, assuming they all readily produce X-ray emission. 
Some fraction of these might be cluster halos whose X-ray origins are due to ICM.
We find independently using the radio galaxy luminosity function that there should also be approximately 10 halos massive enough to host radio AGN that might give rise to X-ray IC emission at some point in their lifetimes. 
This population likely overlaps with the aforementioned population we predict using halo rarity arguments: a cluster halo could host both an ICM signal and an AGN that generates X-ray IC emission. 
On the other hand, some of these halos will not emit X-rays at all at any given time: a halo massive enough to host an ICM will not necessarily have heated sufficiently to emit yet, or an AGN may already have turned off and both the radio and X-ray signals died off at the observed epoch. 
Using the modeled lifetimes of the radio and X-ray components of a lobed AGN, we have conservatively estimated that 4--5 IC ghosts should be visible in the field at $2<z<3$ above $3\sigma$, representing approximately half of all massive halos that should be detectable in the field. 
From this we conclude that at least 40 $\pm$ 17\% of the extended X-ray halos extant in COSMOS at $2<z<3$ should be IC ghosts (see purple line in Figure \ref{fig:halorarity}). 
We cannot accurately predict when the ICM should turn on \citep[e.g.,][]{Chiang2017a}, but by process of elimination we can estimate that the other $\sim$5 halos observable in extended X-rays may arise from the ICM, generously assuming all massive halos have sufficient hot gas to generate ICM emission. 
They may also be IC ghosts given uncertainties in their number densities at this epoch, or these halos may not produce any extended X-ray emission at all.
Our relative assessment of rarity of these phenomena is in line with the predictions from \citet{Mocz2011a}, who show that at $L_X \sim 10^{44}$\,erg/s, the number of IC ghosts is approximately equal to the number of X-ray clusters by $z=2$. 
At face value, this makes either scenario --- ICM or IC ghost --- plausible physical explanations for this X-ray source, with slightly more weight given to the IC ghost description when we also consider the evidence in Section \ref{sec:icghost}.
Related to the actual X-ray observations considered here, we conclude that the proposed protocluster core is most likely associated with an IC ghost instead of an ICM signal from the highest-redshift virialized core. 

\subsection{Rarity of a highly star-forming compact galaxy group}
Since the X-ray detection in question here is itself marginal, it is not this X-ray source in isolation that makes this proposed protocluster core so extraordinary. 
Instead, it is the X-ray signal in addition to its extreme star formation activity in the form of multiple rare star-forming galaxies with demonstrated fast duty cycles (with a total depletion timescale of $\sim$350\,Myr, Section \ref{sec:physstate}). 
We thus use other elements of the observational data to determine how common star-forming galaxy overdensities like in this protocluster core might be in the COSMOS field using the \citet{Laigle2016a} photometric redshift catalog. 

We search for the observed number of overdensities in the catalog using a near-IR selection criterion, which affords us the angular resolution that selecting by submillimeter wavelengths (e.g., SPIRE or \scubaii) would lack and allows us to select for red star-forming galaxies. 
First, we imposed a photometric redshift cut of $z_{phot} < 4$, so that we are selecting for overdensities which may coincide with the X-ray-detectable halos discussed in the previous section. 
We also imposed a magnitude range of $21.5<m_{K_s}<23.4$, which spans from the $3\sigma$ detection limit to the magnitude at which most galaxies at $z>1$ do not exceed \citep{Davidzon2017a}, since we also want to filter out low-redshift overdensities. 

We define the surface density on the sky $\Sigma_N$ by a nearest-neighbor algorithm:

\begin{equation}
\Sigma_N = \frac{N}{\pi r_N^2}
\end{equation} 

\noindent where N is the Nth neighbor and $r_N$ is the distance (in arcseconds) to the Nth-nearest neighbor. 
After determining the average 2D density of $K_s$-band sources in COSMOS, we decided that $N=10$ is a sufficient number of neighbors to determine a spatial overdensity above the field. 
We use a k-d tree to construct the catalog of tenth-nearest neighbors, first in two dimensions (only sky positions) and then in three by including their photometric redshifts (spectroscopic when available). 
A group of galaxies is then declared a 2D overdensity if its tenth nearest neighbor is closer than 50\arcsec\,, which is a distance significance of 2$\sigma$ above the average $r_{10}$ in the catalog.
A 3D overdensity is declared if $\sigma_{\rm z} \leq 0.2$ for the ten sources in question, comparable to the uncertainties in photometric redshift overall. 

Using this simple $K_s$-band magnitude cut, we predict 204 chance alignments along the line of sight that look like overdensities on the sky. 
Because there are so many galaxies in this catalog (35,000), true overdensity signals are likely washed out by outliers, because even the protocluster discussed here is not detected as a 2D overdensity. 
Therefore we impose an additional color cut to select for massive red star-forming galaxies whose neighbors have similar colors, where $J-K_s\geq1.3$ \citep[consistent with the criteria outlined in][]{Wang2016a}. 

The final catalog of 6,000 galaxies contains 16 2D associations of 10 neighboring galaxies and 2 3D associations across the whole COSMOS field, both including the core in question. 
We then searched the \textit{Herschel} SPIRE maps for bright ($>4\sigma$), red ($S_{500}>S_{250}$) counterparts to these associations, as this has frequently been used to select for obscured star formation in galaxy overdensities \cite[e.g.,][]{Clements2014a, Greenslade2017a}. 
Only one candidate overdensity passes this criterion, which is the core discussed here. 

Finally, we calculate the ``probability to exceed" measure from \citet{Harrison2013a}, finding that there is only a 30\% chance that this is the most massive observable structure at $2<z<3$ in COSMOS. 
Since no clear candidates emerge from our selection criteria, they could be found through other selection methods such as pairwise angular correlation functions \citep{Ando2020a} or Ly$\alpha$ tomography \citep{Horowitz2019a}. 

We summarize the probabilities of observing the X-ray detections discussed in Section \ref{sec:prob} and the $K_s$-band galaxy overdensities in Figure \ref{fig:halorarity}. 
We expect approximately 10 halos of similar $z=0$ halo masses to this protocluster core could exist in the COSMOS field if they are detectable via extended X-ray emission. 
Using the COSMOS near-IR catalog, we showed that a comparable number of 2D overdensities within 50\arcsec\, could exist in the same field, although many of these will be chance alignments along the line of sight. 
Both forms of X-ray emission may be more common than the 3D overdensities of red massive galaxies, as we find only 2 3D overdensities across the full field. 
We find only one of these candidates --- this structure -- is also associated with an unusually bright red \herschel\, detection associated with such prodigious star formation (2290\,\sfr), making this structure one of the most unique overdense objects in COSMOS below $z=4$. 

\section{Pointing 2: A normal star-forming filament}\label{sec:pointing2}
Pointing 2 contains one CO-detected source, associated with the galaxy identified in \citet{Casey2015a} as 450.58 at $z=2.464$. 
As seen in Figure \ref{fig:45058}, the source is unresolved, but the spectrum shows a double-peaked CO line.
This is consistent with a rotating disk, though at this spatial resolution it could also indicate two unresolved components of a merger.
Thus, we cannot make a statement about the morphology of this galaxy, but we can place an upper limit on its size of $\sim$40 kpc.
It has a high star formation rate of 380\,\sfr\, but its gas fraction (50\%) is in line with other $z\sim2$ DSFGs of similar gas mass \citep{Scoville2016a}, showing no evidence for enhanced star formation efficiency (see Figure \ref{fig:sfr_ms}).

It is also not detected in CO(3-2) so we report a 2$\sigma$ upper limit of r$_{31} < 0.72$, suggesting the ISM gas in this galaxy is not highly excited and is normal compared to other submillimeter-bright galaxies at similar redshift \citep{Sharon2016a}. 
Interestingly, at least three other LBGs with known redshifts fall within the tuning of this VLA pointing, but they remain undetected in CO, suggesting this filament does not contain other galaxies with particularly rich molecular gas reservoirs (larger than $2.5\times10^{10}$\,\msun). 
All of these galaxies have reported SFRs $<$50\,\sfr, with star-forming gas below the detection limit, so the 450.58 galaxy seems to be part of a less active filament in the parent $z=2.47$ structure. 
We use abundance matching to determine a halo mass since we cannot measure a resolved CO size. 
Using the halo mass relations from \citet{Behroozi2010a}, we estimate a $z=2.5$ halo mass of $\sim$1.7\,$\times$\,10$^{12}$\,\msun, which translates to a $z=0$ halo mass of $1.0\times10^{13}$\,\msun. 

\section{Conclusions}\label{sec:conclusions}
In this paper we have presented the full multiwavelength characterization of eight galaxies detected in molecular gas and associated with a massive galaxy protocluster at $z=2.47$ in two different filaments. 
One galaxy in Pointing 2, 450.58, is isolated along a normal star-forming filament associated with a parent structure at $z=2.47$. 
We showed that this is evidence for constant, steady star formation pathways in normal LBGs within protoclusters.  

Another six galaxies are concentrated within 20\arcsec\, (140 proper kpc), with another associated galaxy, ID5, about 45\arcsec\, away. 
These galaxies, comprising the proposed protocluster core, have a total stellar mass of (1.1 $\pm$ 0.5) $\times10^{12}$\,\msun, a total dust mass of (5.3 $\pm$ 0.5) $\times10^{9}$\,\msun, and a total molecular gas reservoir of (7.8 $\pm$ 0.6)  $\times 10^{11}$\,\msun. 
We find a total SFR of $\approx$3000\,\sfr\, so the core shows no evidence for enhanced star formation efficiency in the cluster environment relative to the field at matched redshift.

We presented two hypotheses to explain the observed characteristics of the protocluster core: Hypothesis A, in which this structure represents the highest-redshift virialized cluster core, and Hypothesis B, in which the galaxies are separated along the line of sight and the close 20\arcsec\, association is a mix of projection and interaction. 
This structure raises important questions about the ``inside-out" nature of collapse within protoclusters -- if Hypothesis A indeed holds, protoclusters could form embedded, virialized cores prior to the collapse of the larger structure. 
An alternative scenario would be that the largest extended structures collapse all at once, while the X-ray ICM signal follows later. 

We summarize the results of the two hypotheses here:

\begin{itemize}
    \item We revisited the coincident X-ray detection that has previously been claimed as the primary evidence for the highest-redshift cluster core, consistent with our Hypothesis A. 
    Under Hypothesis B, we presented evidence for the X-ray emission arising from an Inverse Compton ghost associated with a galaxy embedded within the filament. 
    Based on the luminosity and light profile of the marginal (3$\sigma$) \textit{Chandra/XMM} detection, we showed that the emission is more likely to be associated with a recently extinguished radio galaxy.
    
     \item We assessed the relative rareness of this protocluster core, concluding that roughly 10 X-ray-selected protoclusters may be visible out to $z=4$ provided that all produce sufficient X-ray emission, which may not be the case. 
     We estimate using halo rarity and AGN lifetime arguments that at least 40 $\pm$ 17\% of halos hosting extended X-ray emission at $z<4$ contain IC ghosts. 
     We conclude that the IC ghost explanation is more likely than X-ray emission from ICM.
    
    \item The galaxies show no evidence for enhanced star formation efficiency and their gas fractions are similar to those found for main sequence galaxies in the field. 
    Despite the relatively poor resolution of the VLA data, the aggregate 3D CO distribution has a complex morphology, making interactions between galaxies possible.
    We showed through dynamical arguments that the galaxies are unlikely to reside in a single virialized halo as in Hypothesis A.
    
    \item We calculated halo masses for the $z=2.5$ core system using several methods under the two hypotheses and find that regardless of the current configuration these galaxies will likely become gravitationally bound to the parent $z=2.47$ structure and could possibly form a BCG with an equivalent total halo mass of 2--9 $\times 10^{14}$\,\msun\, at $z=0$. 
    
    \item Using the observed catalogs of K-band-selected galaxies, we showed there are probably fewer than 15 structures within COSMOS showing similar overdensities in star-forming galaxies, although there is only a 30\% chance that the $z=2.5$ core is the most massive overdensity in COSMOS. 
    
\end{itemize}

Therefore, we suggest that previous claims that this is the highest redshift cluster is less secure than previously thought. 
We conclude that the richly star-forming core in Pointing 1 is more consistent with Hypothesis B, where this structure is an extended filament along the line of sight which may host a faint IC ghost and is still actively undergoing mergers and gas interactions. 
This core represents an extraordinary laboratory to investigate the physical processes of protocluster evolution, as we have shown the diversity of possible protocluster configurations during the peak of cosmic star formation in the Universe. 

\vspace{-0.75cm}
\section*{Acknowledgements}
We thank the anonymous referee for their insightful suggestions. 
JBC and CMC thank the National Science Foundation for support through grants AST-1714528 and AST-1814034, and additionally the University of Texas at Austin College of Natural Sciences. 
In addition, CMC acknowledges support from the Research Corporation for Science Advancement from a 2019 Cottrell Scholar Award sponsored by IF/THEN, an initiative of Lyda Hill Philanthropies. 

This paper makes use of the data from the Karl G. Jansky Very Large Array. 
The National Radio Astronomy Observatory is a facility of the National Science Foundation operated under cooperative agreement by Associated Universities, Inc. 
This paper also makes use of the following ALMA data: ADS/JAO.ALMA\#2016.1.00646.S. ALMA is a partnership of ESO (representing its member states), NSF (USA) and NINS (Japan), together with NRC (Canada) and NSC and ASIAA (Taiwan) and KASI (Republic of Korea), in co-operation with the Republic of Chile. The Joint ALMA Observatory is operated by ESO, AUI/NRAO and NAOJ. 

The Flatiron Institute is supported by the Simons Foundation.

H.D. acknowledges financial support from the Spanish Ministry of Science, Innovation and Universities (MICIU) under the 2014 Ram\'{o}n y Cajal program RYC-2014-15686 and AYA2017-84061-P, the later one co-financed by FEDER (European Regional Development Funds).

\software{magphys (da Cunha et al. 2008, 2015), Hyper-z SED (Bolzonella et al. 2000), CASA (McMullin et al. 2007), Astropy \citep{Astropy18}, Matplotlib \citep{Hunter2007a}}

\nocite{*}
\bibliography{main}

\appendix
\section{Details of VLA 33 GHz Data Analysis}\label{app:appendix}
\subsection{Source Extraction Procedure}\label{app:extraction}

We extract spectra in two ways from the each VLA 33\,GHz cube: (1) a single-pixel extraction at the position of the peak flux in moment-0 maps, with the channels centered on the expected redshift of the source, and (2) extraction at the same position but with a custom aperture, where all adjacent pixels have SNR$>$3 (typically the size of 1--2 VLA beams), taking care to place the apertures such that we do not double-count flux between nearby sources.

If the sources were resolved, we would expect there would be a significant excess in the integrated flux density measured in a large aperture versus the brightest pixel, whereas an unresolved source would show little variation in the integrated flux density regardless of the aperture size (provided there is not a significant velocity gradient across the source).

\begin{figure}
\includegraphics[width=0.5\columnwidth]{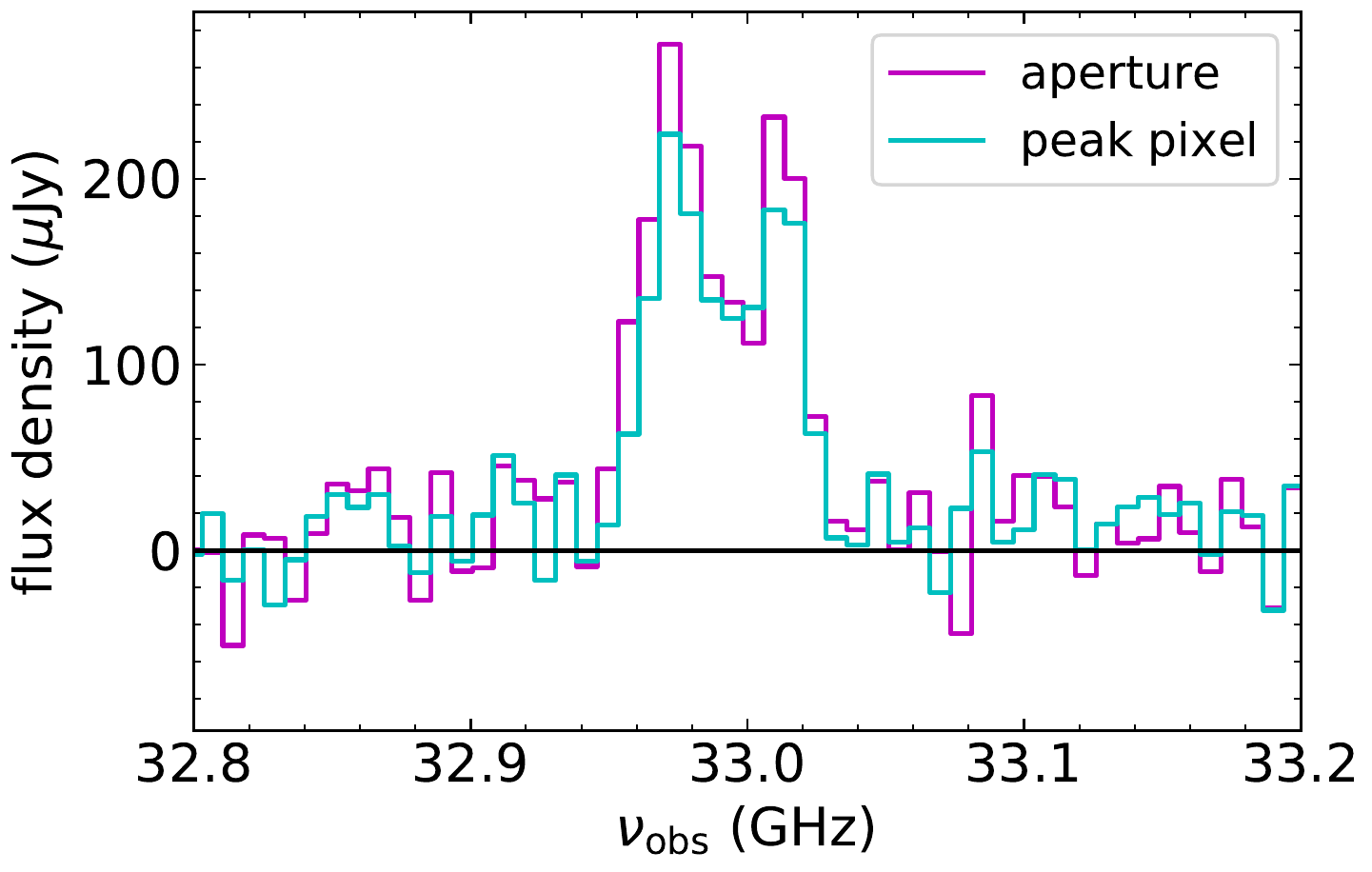}
\includegraphics[width=0.45\columnwidth]{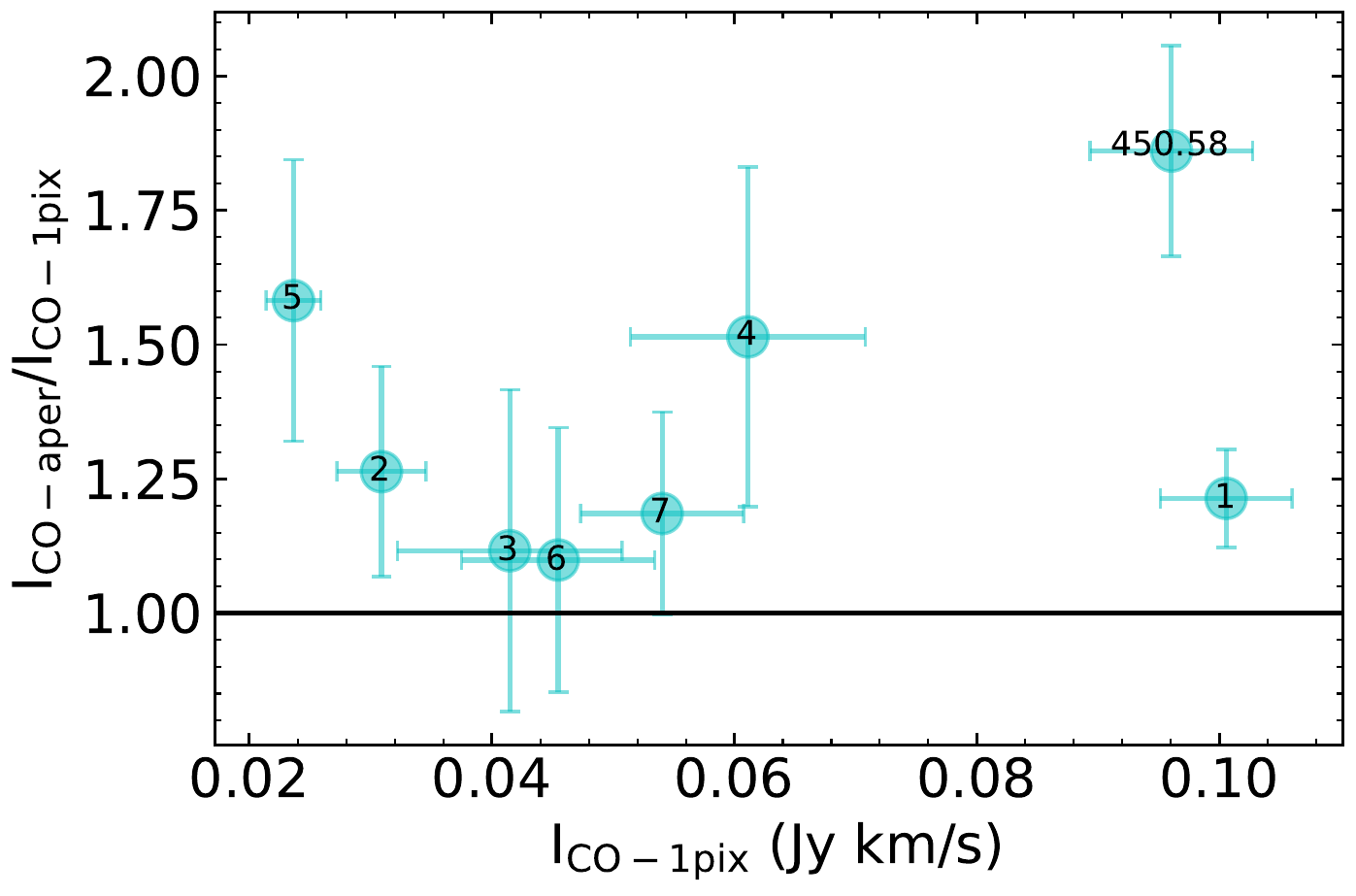}
\caption{\textit{Top:} Example spectrum of ID1, the brightest CO(1-0) source, extracted in two ways: a single-pixel spectrum taken from the brightest pixel (cyan) and integrated from a custom aperture (magenta). 
Both the total flux and the shape of the line profile are preserved in both methods, showing that these sources are unresolved. 
\textit{Bottom:} Ratio of integrated flux density measured from the two methods of spectrum extraction versus the single-pixel extraction intensity, with the numbers corresponding to the ID numbers in Table \ref{table:co}. 
The total flux density is consistent with unity, or at most a factor of 2 difference, showing that these sources are at most only marginally resolved.}
\label{fig:aperture}
\end{figure}

In the top panel of Figure \ref{fig:aperture} we show an example spectrum of the brightest source (ID1) extracted with each of these two methods and show that the line profiles do not differ between an aperture and the brightest pixel, which we expect since the sources have low enough S/N that we would not be sensitive to velocity gradients across the source.
In the bottom panel of Figure \ref{fig:aperture} we show that most sources are unresolved or marginally resolved by the ratio of the line intensity $I_{\rm CO}$ extracted in a custom aperture to that measured from the single brightest pixel. 
Most sources are roughly consistent with unity indicating they are point-like sources, but several of the brighter sources (ID1, ID4, and 450.58) appear marginally resolved, where we measure 25--75\% more flux within a beam-sized aperture. 

Another way to test whether sources are resolved would be to fit a 2D Gaussian model to each source and compare the estimated size with the size of the VLA beam.
We tried this using the CASA task \textsc{imfit}, but we encounter catastrophic failures or unphysically large fit sizes due to the crowding of the sources and likely decorrelation of the signal at high observing frequencies.
We additionally re-imaged the cubes using \textsc{uvtaper} to degrade the spatial resolution and test whether we have resolved out any flux.
Regardless of the spatial resolution, we find total aperture fluxes consistent within 10\% and a consistent inability to fit individual 2D Gaussian models to the sources.
Performing the same analysis in the $uv$-plane or on cleaned data does not change the resolution results. 

\subsection{Calculating Spectral Line Moments for CO(1-0)}\label{app:moments}

We calculate the CO line properties by directly measuring the spectral line moments, following the procedures of \citet{Bothwell2013a}.
The CO redshift is determined by the first moment, which is the intensity-weighted line center.
The intensity-weighted second moment determines the line width, which is defined as:

\begin{equation}
    s_{\nu} = \frac{\int(\nu-\bar{\nu})^2 I_{\nu}\, d\nu}{\int I_{\nu}\, d\nu}
\end{equation}

\noindent such that the Gaussian full width at half maximum (FWHM) is 2$\sqrt{2\, \rm ln2}$\,$s_{\nu}$. 

The calculation of spectral moments is sensitive to the noise in the spectrum, so we run a Monte Carlo simulation 2000 times, in which Gaussian noise ($\sigma$ equal to the rms of the line-free spectrum) is artificially added to the spectral data and the line moments recalculated.
(Artificially adding noise results in a reported error that is a factor of $2^{-1/2}$ higher than the true noise, but this is accounted for in the final results.)
In our Monte Carlo procedure, we restrict the line search to a radius of 100\,\kms\, from the prior guess for the redshifts. 
The peak, center, and FWHM are determined by the medians of the results of this procedure, with the errors as the standard deviations in these distributions.
We additionally multiply the spectrum by $-$1 and repeat the procedure to make sure there are no negative $3\sigma$ ``sources" in the spectrum, of which we find none.

Finally, the integrated line intensity $I_{\rm CO}$ is formally the velocity-integrated line flux along $\pm2\sigma$ from the line center, but we simply sum the flux within the range where the signal remains positive in adjacent channels. 

\subsection{Measuring Gas Masses}\label{app:aco}
We use the CO(1-0) line measurements to calculate molecular gas mass, using the usual factor \alphaco\, to convert from line luminosity to total molecular gas mass, but the appropriate value of \alphaco\, to use can be ambiguous.
The Galactic value of \alphaco\, = 4\,\acounits\, 
was calibrated using observations of individually resolved giant molecular clouds (GMCs) within the Milky Way and nearby spiral galaxies, while local luminous infrared galaxies (LIRGS) were presumed to have much lower conversion factors of \alphaco\, $\sim$ 0.8\,\acounits, with CO emission stemming from a warm, diffuse interstellar medium rather than virialized GMCs \citep*[see review of][]{Bolatto2013a}. 
While \alphaco\, is likely dependent on metallicity and the thermal state of molecular gas clouds, a value of \alphaco\,=1\,\acounits\, is typically used for unresolved, compact submillimeter galaxies \citep*{Bolatto2013a, Carilli2013b} at high redshift. 
If galaxies are dominated by dense, nuclear starbursts, then a low value of \alphaco\, is appropriate, which has been a valid assumption especially in the case where higher \alphaco\, values yield gas masses in excess of the dynamical masses of galaxies.

However, \citet{Scoville2016a} broadly argues for a higher value of \alphaco\, in DSFGs, especially in cases where the densest sites of star formation -- where there is significant dust heating -- are not spatially resolved. 
Instead, they advocate for using a conversion factor that is globally averaged, where galaxies are not dominated by these hot starbursting regions. 
In this case the authors advocate for the Galactic value of \alphaco\, which is \alphaco\,=\,4.5\,\acounits\, for the mass of $H_2$, or \alphaco\,=\,6.5\,\acounits\, if assuming a factor of 1.36 for the contribution of helium. 
We adopt this Galactic value of \alphaco\,=\,6.5\,\acounits\, here to indicate the total expected molecular gas mass.



\end{document}